\documentclass[a4paper,fleqn,usenatbib]{mnras}
\usepackage{graphicx}
\usepackage[dvipsnames]{xcolor}
\usepackage{siunitx}
\usepackage{verbatim}
\usepackage[english]{babel}
\usepackage[T1]{fontenc}
\usepackage{ae,aecompl}

\usepackage{subfig} 
\usepackage{float}

\usepackage{color}
\usepackage{multirow}
\usepackage{lscape}
\usepackage{longtable}
\usepackage{caption}

\newcommand{\xmm}{{\em XMM-Newton}}
\newcommand{\sw}{{\em Swift}}
\newcommand{\chan}{{\em Chandra}}

\title[Observations of candidate milli-second pulsars]{A multi-wavelength investigation of candidate milli-second pulsars in unassociated $\gamma$-ray sources}

\author[D. Salvetti et al.]{
\parbox{\textwidth}{
D. Salvetti$^{1}$\thanks{E-mail: salvetti@iasf-milano.inaf.it},
R. P. Mignani$^{1,2}$, 
A. De Luca$^{1,3}$,
M. Marelli$^{1}$,
C. Pallanca$^4$,
A. A. Breeveld$^5$,
P. H\"usemann$^6$,
A. Belfiore$^1$,
W. Becker$^{6,7}$, and
J. Greiner$^6$
} 
\\ \\
$^{1}$ INAF - Istituto di Astrofisica Spaziale e Fisica Cosmica Milano, via E. Bassini 15, 20133, Milano, Italy\\
$^{2}$ Janusz Gil Institute of Astronomy, University of Zielona G\'ora, Lubuska 2, 65-265, Zielona G\'ora, Poland \\
$^{3}$ INFN - Istituto Nazionale di Fisica Nucleare, sezione di Pavia, via A. Bassi 6, 27100, Pavia, Italy \\
$^4$ Dipartimento di Fisica e Astronomia, Universit\`a degli Studi di Bologna, Viale Berti Pichat 6-2, I-40127, Bologna, Italy \\
$^5$ Mullard Space Science Laboratory, University College London, Holmbury St. Mary, Dorking, Surrey, RH5 6NT, UK \\
$^6$ Max-Planck Institut f\"ur extraterrestrische Physik, Giessenbachstrasse 1, 85741 Garching, Germany\\
$^7$Max-Planck Institut f\"ur Radioastronomie, Auf dem H\"ugel 69, 53121 Bonn, Germany
}

\date{Accepted XXX. Received YYY; in original form ZZZ}
\pubyear{2016}

\begin{document}
\label{firstpage}
\pagerange{\pageref{firstpage}--\pageref{lastpage}}
\maketitle

\begin{abstract}
About one third of the 3033 $\gamma$-ray sources in the Third {\em Fermi}-LAT Gamma-ray Source Catalogue (3FGL) are unidentified and do not have even a tentative association with a known object, hence they are defined as {\em unassociated}. Among Galactic $\gamma$-ray sources, pulsars represent the largest class, with over 200 identifications to date. About one third of them are milli-second pulsars (MSPs) in binary systems. Therefore, it is plausible that a sizeable fraction of the unassociated Galactic $\gamma$-ray sources belong to this class. We collected X-ray and optical observations of the fields of twelve unassociated {\em Fermi} sources that have been classified as likely MSPs according to statistical classification techniques. To find observational support for the proposed classification, we looked for periodic modulations of the X-ray and optical flux of these sources, which could be associated with the orbital period of a MSP in a tight binary system. Four of the observed sources were identified as binary MSPs, or proposed as high-confidence candidates, while this work was in progress. For these sources, we present the results of our follow-up investigations, whereas for the others we present possible evidence of new MSP identifications. In particular, we discuss the case of 3FGL\, J0744.1$-$2523 that we proposed as a possible binary MSP based upon the preliminary detection of a 0.115 d periodicity in the flux of its candidate optical counterpart. We also found very marginal evidence of periodicity in the candidate optical counterpart to 3FGL\, J0802.3$-$5610, at a period of 0.4159 d, which needs to be confirmed by further observations.
\end{abstract}


\section{Introduction}

Observations carried out with the {\em Fermi} Gamma-ray Space Telescope have produced an unprecedented harvest of $\gamma$-ray sources thanks to the improved performance of its Large Area Telescope (LAT; Atwood et al.\ 2009) with respect to its predecessor, the Energetic Gamma Ray Experiment Telescope (EGRET; Thompson et al.\ 1993), which flew on the {\em Compton} Gamma-ray Observatory ({\em CGRO}). The most recent catalogue of $\gamma$-ray sources detected by the LAT, the Third {\em Fermi}-LAT Gamma-ray Source Catalogue (3FGL; Acero et al.\ 2015), based on four years of data (2008--2012), contains 3033 sources, a factor of ten more than detected by the {\em CGRO} in a similar time span (1991--1995; Hartman et al.\ 1999). The identification of these sources, however, is in many cases a challenging task owing to the still relatively large $\gamma$-ray error regions and to the problems of identifying a signature that can unambiguously reveal the source nature. In the case of $\gamma$-ray pulsars, the detection of pulsations represents such a signature, and over 200 $\gamma$-ray sources have been identified in this way\footnote{For the public list of $\gamma$-ray pulsars detected by the LAT, see {\tt https://confluence.slac.stanford.edu/x/5Jl6Bg}} (see Grenier \& Harding 2015 for a recent review on $\gamma$-ray pulsars), which are either radio loud (RL), i.e. also detected as radio pulsars, or radio quiet (RQ), i.e. undetected in radio in spite of deep searches. In the case of Active Galactic Nuclei (AGN), the second most numerous class of identified $\gamma$-ray sources, the characteristic signature is represented by the observation of long-term variability, possibly correlated with a similar activity in the radio band. The number of sources in the 3FGL that have been identified through a characteristic signature represents, however, a tiny minority. Indeed, while 1785 $\gamma$-ray sources in the 3FGL have been at least {\em associated} with objects in master catalogues based upon positional coincidence (mainly AGNs), another 1010 sources have no association at all, hence referred to as {\em unassociated}. Needless to say, ascertaining the nature of these sources has far reaching implications in the understanding of the $\gamma$-ray emission of our Galaxy. 

Since $\gamma$-ray pulsars represent  the most numerous class of identified Galactic $\gamma$-ray sources, it comes as natural that many efforts have been focused on the search for these objects. Apart from relatively young ($\la 1$ Myr old) $\gamma$-ray pulsars, the LAT has identified a surprisingly large number of milli-second pulsars (MSPs). These are old ($\ga$ 1 Gyr) pulsars, much less energetic than the younger ones, with a rotational energy loss rate of $\dot{E}_{\rm rot} \sim 10^{33}$ erg s$^{-1}$, which owe their fast rotation periods (a few milliseconds, hence the name) to the spin up following a phase of matter accretion from a donor companion star. Not surprisingly, out of the 93 $\gamma$-ray MSPs, 73 are found in binary systems. They have usually either a white dwarf (WD) or a low-mass main sequence (MS) companion, such as the so-called Black Widows (BWs) and Redbacks (RBs), which have companion masses $M_{\rm com} < 0.1 M_{\odot}$ and $M_{\rm com} \sim 0.1-0.4 M_{\odot}$, respectively (Roberts 2013). The low companion masses explain the short orbital periods ($P_{\rm orb} < 1$ d) of these systems and fit the scenario where the companions are ablated by irradiation from the pulsar wind, a process that would explain the existence of solitary MSPs. This scenario has recently received observational support from the identification of some RB systems that alternates between states of accretion from the companion, when the X-ray emission increases and the radio emission is temporarily quenched by the accreting material, and  states of no accretion, where the X-ray emission decreases and the radio emission reactivates (e.g., Patruno et al.\ 2014), hence dubbed ``transitional''. 

Many BWs and RBs have been found to lurk in unassociated $\gamma$-ray sources detected by {\em Fermi}, and their number has now increased by at least a factor of five. The difficulties in finding such systems in the traditional radio channel, i.e. by detecting them as radio MSPs, is intrinsic to the very nature of such systems. Even in the non-accretion phases, a considerable amount of plasma is present in the intra-binary environment, produced by the irradiation of the companion star surface, which causes eclipses and delays of the radio emission from the MSP. This makes BWs and RBs very elusive targets in radio and their identification requires observations at different wavelengths, e.g. in the X-rays and in the optical. In a number of cases, these observations have been instrumental in pinpointing candidate BW/RB systems in unassociated {\em Fermi} sources and paved the way for the discovery of radio/$\gamma$-ray pulsations. Since BWs and RBs are characterised by short orbital periods, the most successful strategy consists of searching for $<1$ day periodic modulations of the optical flux of the putative companion star. These are produced by irradiation and tidal distortion effects that affect the star brightness when it is seen at different orbital phases. This strategy was applied to pinpoint the MSPs associated with the $\gamma$-ray sources 1FGL\, J1311.7$-$3429 (Romani et al.\ 2012) and 1FGL\, J2339.7$-$0531 (Kong et al.\ 2012), and other likely MSP candidates associated with 2FGL\, J1653.6$-$0159 (Romani et al.\, 2014), 2FGL\, J0523.3$-$2530 (Strader et al.\ 2014), 3FGL\, J2039.6$-$5618 (Salvetti et al.\ 2015; Romani 2015) and 3FGL\, J0212.1+5320 (Li et al.\ 2016).

How many unassociated $\gamma$-ray sources, especially at high Galactic latitude where MSPs have time to migrate on Gyr time scales thanks to their proper motions, can be identified as either BWs or RBs, is an open issue. To help addressing this issue, back in 2014 we started a project to search for BWs/RBs in a selected sample of unassociated {\em Fermi}-LAT sources based on the detection of optical modulations with a few hour periods from the putative MSP companions. A preliminary account of our project, with the discussion of early results for the most interesting sources, is given in Mignani et al.\ (2016). Our manuscript is structured as follows: in Sectn. 2, we describe the selection of the MSP candidates and the multi-wavelength observations. In Sectn. 3, we present the results, and discuss their implications in Sectn. 4.

\section{Observations and data reduction}

\subsection{Candidate selection}

As a first step, we selected a starting sample of MSP candidates from the unassociated {\em Fermi}-LAT sources. Since our project started in 2014, we originally selected these sources from the 2FGL catalogue (Nolan et al.\ 2012), which was the reference catalogue of {\em Fermi}-LAT sources available back then. All the sources selected are also listed in the 3FGL catalogue (Acero et al.\ 2015), which from now on we assume as a reference. The candidate selection was based on their $\gamma$-ray spectral and temporal characteristics and was implemented through the results of an artificial neural network classification code, fully described in Salvetti (2016). Our candidate selection method agrees with that of Saz Parkinson et al.\ (2016), which also classified all our candidates as MSPs. This starting sample was the basis of our multi-wavelength investigations aimed at confirming the source classification on firm observational evidence.

As a second step, we narrowed down the candidate selection to those $\gamma$-ray sources with small $\gamma$-ray error ellipses ($r95 \le 0\fdg1$) and with X-ray coverage from either {\em XMM-Newton}, {\em Chandra} or {\em Swift}. This is because MSPs are also X-ray sources, with the X-ray emission produced from the pulsar magnetosphere (or hot polar caps on the pulsar surface) and/or from the intra-binary shock (Roberts 2013). 

Finally, we selected candidates for which multi-epoch photometry measurements were available from public optical sky surveys and/or for which we obtained follow-up imaging observations with ground-based optical facilities (see Sectn.\ 2.3 for details). In this way, we selected a sample of 12 MSP candidates, including the suspected RB candidates 3FGL\, J0523.3$-$2528, J1653.6$-$0158, and J2039.6$-$5618 (Table~\ref{opt}). We refer to Acero et al.\ (2015) for the characterisation of the $\gamma$-ray properties of these sources and to Saz Parkinson et al.\ (2016) for the description of their classification as MSP candidates. An account of the available X-ray and optical observations is given in the next two sections and is summarised in Table~\ref{opt}.

\subsection{X-ray observations}

All our targets have an adequate X-ray coverage of the LAT error boxes with either {\em Swift} (Burrows et al.\ 2005) or {\em XMM-Newton} (Str\"uder et al.\ 2001; Turner et al.\ 2001). All of them have been observed by \sw\ as  part of a systematic survey of the $\gamma$-ray error boxes of the unassociated {\em Fermi}-LAT sources (Stroh \& Falcone 2013). In all cases, the \xmm\ observations have been executed as pointed follow-ups of the LAT sources. Only 3FGL\, J1653.6$-$0158 (Cheung et al.\ 2012) has been observed with \chan\ (Garmire et al.\ 2003).

We reduced and analysed the \xmm\ data through the most recent release of the \xmm\ Science Analysis Software (\emph{SAS}) v15.0. We performed a standard data processing, using the {\tt epproc} and {\tt emproc} tools, and screening for high particle background time interval following Salvetti et al.\ (2015). For the \chan\ data analysis we used the Chandra Interactive Analysis of Observation (\emph{CIAO}) software version 4.8. We re-calibrated event data by using the {\tt chandra\_repro} tool. \sw\ data were processed and filtered with standard procedures and quality cuts\footnote{More detail in: {\tt http://swift.gsfc.nasa.gov/docs/swift/analysis/}} using FTOOLS tasks in the {\sc HEASOFT} software package v6.19 and the
calibration files in the latest Calibration Database release.

\subsection{Optical observations}

 Ten sources in our sample were observed as part of the Catalina Sky Survey\footnote{{\tt http://www.lpl.arizona.edu/css/}}, a program of three optical sky surveys covering the whole sky $15^{\circ}$ above and below the Galactic plane (Drake et al.\ 2009). In the northern hemisphere, the Catalina Sky Survey is carried out with two wide-field telescopes: the 0.7m Schmidt at the Mount Bigelow observatory (Arizona), which produces the Catalina Schmidt Survey (CSS), and the 1.5m reflector at Mount Lemmon observatory (Arizona), which produces the Mount Lemmon Survey (MLS). In the southern hemisphere, the 0.5m Schmidt telescope (Siding Spring, Australia) was retired at the end of 2013 and only archival observations produced by the Siding Spring Survey (SSS) are available. The Catalina Sky Survey is carried out with a cadence of days to months down to limiting magnitudes $V\sim 19.5$--21.5 per pass. 

We carried out dedicated follow-up observations of seven of the MSP candidates in our sample with the Gamma-Ray Burst Optical/Near-Infrared Detector (GROND; Greiner et al.\ 2008) at the MPI/ESO 2.2m telescope on La Silla (Chile). For five of these candidates we have also multi-epoch data from the Catalina Sky Survey. We carried out simultaneous observations in the optical $g',r',i',z'$ and near-infrared $J,H,K$ bands repeated with a regular cadence on two or more consecutive nights. The observations were split into sequences of four 115 s dithered exposures in the optical and forty eight 10 s dithered exposures in the near-IR. To cope with the right ascension distribution of our targets, the observations were divided in two runs, in August 2014 and February 2015. The data were processed and calibrated using the GROND pipeline (Kr\"uhler et al.\ 2008; Yoldas et al.\ 2008). The astrometry calibration was calculated from single exposures by comparison with  stars selected from the USNO-B1.0 catalogue (Monet et al.\ 2003) in the optical bands and the 2MASS catalogue (Skrutskie et al.\ 2006) in the near-IR bands, yielding an accuracy of 0\farcs3 with respect to the chosen reference frame. The photometric calibration in the optical was computed using close-by fields from the Sloan Digital Sky Survey (York et al.\ 2000), whereas in the near-IR it was computed using 2MASS stars in the GROND field of view. The accuracy of the absolute photometry calibration was 0.02 magnitudes in the g$'$, r$'$, i$'$, z$'$ bands, 0.03 magnitudes in J and H, and 0.05 in the K$_s$ band. See H\"usemann (2015) for details on the GROND data processing and calibration. A separate automated variability analysis was then executed on the photometry result files of the standard GROND pipeline for all sources in the field of view using a code developed at MPE, based on a standard deviation analysis of the optical light curve (H\"usemann 2015) and with a threshold $\sigma_{\rm th}=0.05$. In parallel, we ran an independent periodicity search using the Lomb-Scargle algorithm (Zechmeister \& Kuerted 2009), exploring candidate periods above 0.1 d, to include the period range characteristic of BW and RB systems. We then inspected candidate periodicities through a standard folded light curve analysis.

\subsection{X-ray/optical analysis}

As a first step, we looked for candidate X-ray counterparts to the LAT sources that had not yet been firmly identified in the X-rays, i.e. all sources in Table~\ref{opt} except the three RB candidates. We performed a standard data analysis and source detection in the 0.3--10 keV energy band of the \xmm-EPIC, \chan-ACIS and \sw-XRT observations (e.g., Salvetti et al.\ 2015 and Marelli et al.\ 2015). We focused on the X-ray sources detected at a significance $\ga3\sigma$ inside, or close to, the 95\% confidence 3FGL error ellipse to search for the possible counterparts to each $\gamma$-ray source. For each of these X-ray sources we performed a spectral and timing analysis using \emph{XSPEC} v12.9 and \emph{XRONOS} v5.21, respectively. We extracted X-ray fluxes by fitting the spectra with a power-law (PL) model using either a $\chi^2$ or the C-statistic (Cash 1979) in the case of low counts ($<$ 100 photons). For sources characterised by low statistics, we fixed the column density to the value of the Galactic $N_{\rm H}$ integrated along the line of sight (Dickley \& Lockman 1990) and, if necessary, the X-ray PL photon index ($\Gamma_{\rm X}$) to 2. All quoted uncertainties on the spectral parameters are reported at the 1$\sigma$ confidence level. For all sources we computed the corresponding $\gamma$--to--X-ray flux ratio. As reported in Marelli et al.\ (2015), this could still give important information on the nature of the source. For all sources, we also generated background-subtracted X-ray light curves with at least 25 counts per time bin in order to evaluate the variability significance through a $\chi^2$ test. Since BWs and RBs are characterised by an X-ray flux modulation with a period smaller than 1 day, which is associated with emission from the intra-binary shocks (Salvetti et al.\ 2015), we searched for periodic modulations in the barycentred data using the standard power spectrum analysis. Finally, for all observations, we computed the $3 \sigma$ X-ray detection limit based on the measured signal--to--noise ratio, assuming a PL spectrum with $\Gamma_X=2$ and the integrated Galactic $N_{\rm H}$.

As a second step, we looked for variable optical counterparts to the detected X-ray sources in either the Catalina Sky Survey or in the GROND data (or both). Therefore, our strategy is tailored to the identification of binary MSPs. Only when we found a potentially interesting counterpart, i.e. with clear evidence of periodic flux modulation, we also exploited multi-band observations in optical, ultraviolet, and infrared archives for a better counterpart characterisation. We note that it was not possible to use the Catalina Surveys Periodic Variable Star Catalogue (Drake et al.\ 2014) to carry out a systematic search for variable sources across the entire 3FGL error ellipses of our $\gamma$-ray sources since this catalogue only covers the declination region $-22^{\circ} < \delta < 65^{\circ}$ and all our targets, with the only exception of 3FGL\, J1625.1$-$0021, J1630.2$+$3733, and J1653.6$-$0158, are south of it. 

 \begin{table*}
\begin{center}
\caption{Candidate MSPs from unassociated {\em Fermi}-LAT sources discussed in this work. Coordinates and size of the 95\% semi-major axis of the $\gamma$-ray error ellipse (r95) are taken from the 3FGL catalogue (Acero et al.\ 2015). Sources are sorted in RA. The multi-wavelength observations used in this work are summarised in columns five and six.}
\label{opt}
\begin{tabular}{ccccccc} \hline
\multirow{2}{*}{3FGL Name} & RA & DEC & r95 & \multirow{2}{*}{X-ray observations} & \multirow{2}{*}{Optical observations} & \multirow{2}{*}{Notes (References)}\\
 &$^{(hh ~mm ~ss)}$ &  $^{(\circ ~' ~'')}$ &  ($^{\circ}$) & & \\ \hline
\vspace{-3mm}\\
       J0523.3$-$2528         & 05 23 21.4	& $-$25 28 35	& 0.04   & \sw$\dagger$ & Catalina$\dagger$, GROND & Candidate RB (1)\\			
       J0744.1$-$2523		& 07 44 10.7	& $-$25 23 58	& 0.05   & \sw & GROND & \\
       J0802.3$-$5610		& 08 02 19.9	& $-$56 10 08	& 0.10   & \sw, \xmm & Catalina &\\
       J1035.7$-$6720		& 10 35 42.2	& $-$67 20 01	& 0.04   & \sw, \xmm & GROND & Pulsar (2) \\
       J1119.9$-$2204		& 11 19 56.3	& $-$22 04 02	& 0.04      & \sw $\ddagger$, \xmm\ & Catalina$\ddagger$, GROND &\\
        J1539.2$-$3324	& 15 39 17.6	& $-$33 24 51	& 0.04          & \sw$\ddagger$ & Catalina, GROND & \\
       J1625.1$-$0021		& 16 25 07.1	& $-$00 21 31  	& 0.04    & \sw, \xmm$\ddagger$ & Catalina, GROND & \\
       J1630.2$+$3733	& 16 30 12.8	& $+$37 33 44	& 0.07    & \sw & Catalina & Binary MSP (3) \\
       J1653.6$-$0158		& 16 53 40.6	& $-$01 58 48	& 0.04   & \sw, \chan$\ddagger$  & Catalina$\dagger$ &  Candidate RB (4)\\
       J1744.1$-$7619		& 17 44 10.8	& $-$76 19 43	& 0.03   & \sw, \xmm$\ddagger$ & Catalina & Pulsar (2) \\
       J2039.6$-$5618		& 20 39 40.3	& $-$56 18 44	& 0.04   & \sw$\star$, \xmm$\star$ & Catalina$\star$, GROND$\star$ &  Candidate RB (5,6) \\		
       J2112.5$-$3044		& 21 12 34.7	& $-$30 44 04	& 0.04   & \sw, \xmm & Catalina & \\ 
\vspace{-3mm}\\\hline
\vspace{-2mm}\\
 \multicolumn{7}{l}{\footnotesize{
 List of references: (1) Strader et al.\ (2014), (2) Clark et al.\  (2016); (3) Sanpa-Arsa et al., in prep.; (4) Romani et al.\ (2014);}}\\
\multicolumn{7}{l}{\footnotesize{
(5) Salvetti et al.\ (2015); (6) Romani (2015)  }}\\
 \multicolumn{7}{l}{\footnotesize{Data analysis published in $\dagger$Strader et al.\ (2014); $\ddagger$ Hui et al.\  (2015); $\star$ Salvetti et al.\  (2015); Romani (2015)}}\\
 \end{tabular}
\end{center}
\end{table*}

\section{Results}

\begin{figure}
\begin{center}
{\includegraphics[width=5.1cm,angle=270,clip=true]{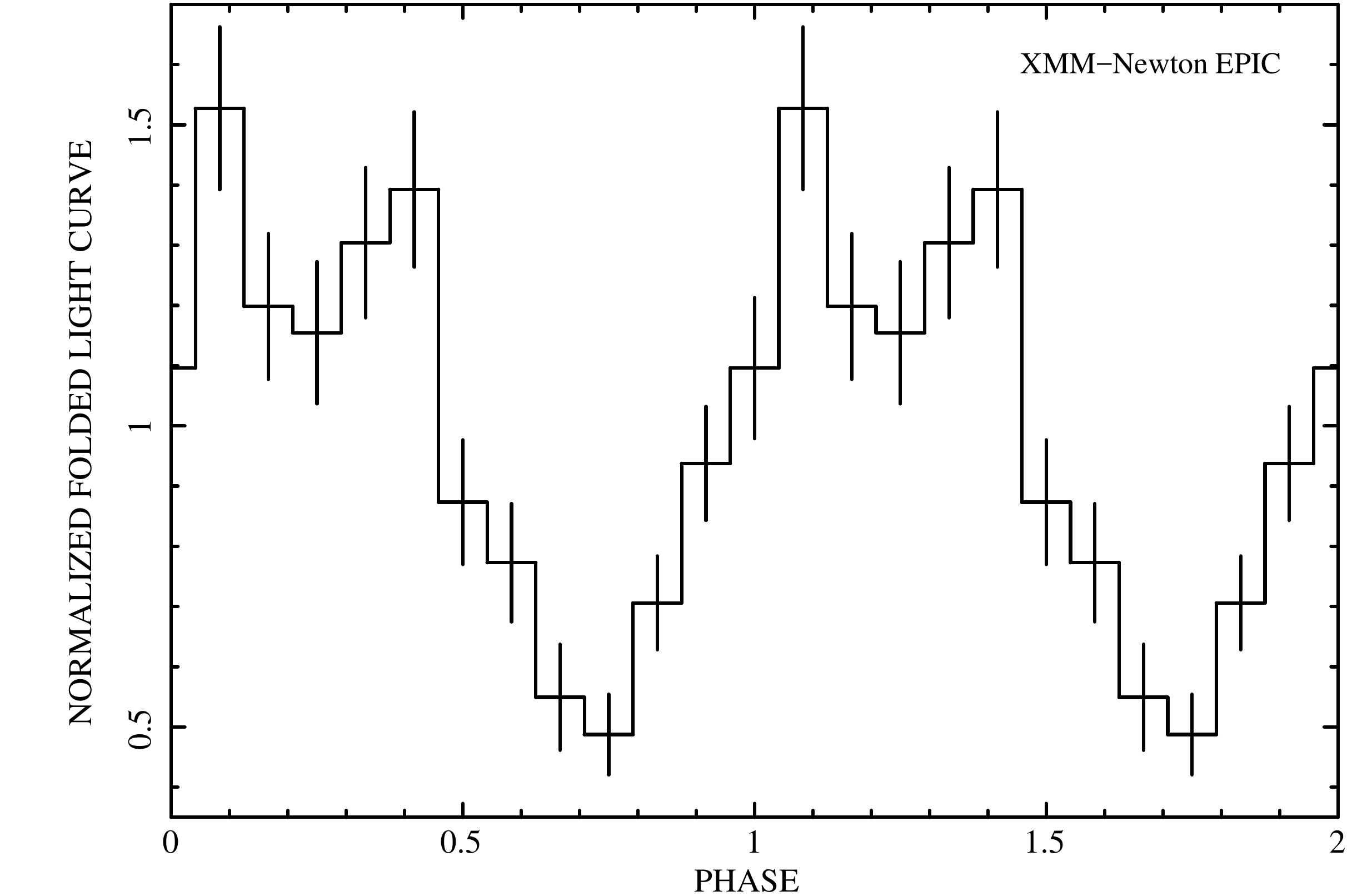}\vspace{.4cm}}
{\includegraphics[width=7.8cm, angle=0,clip=true]{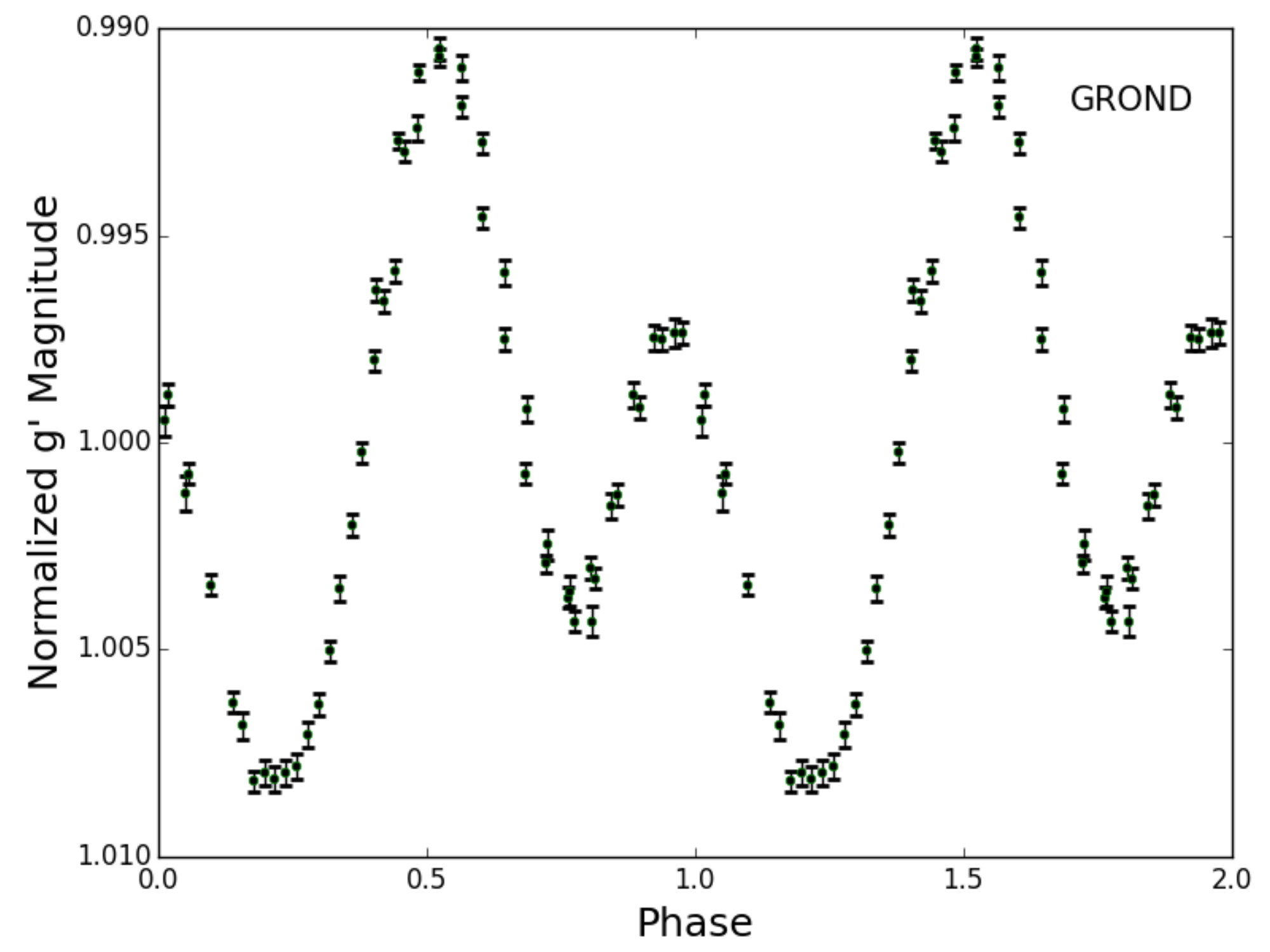}}
{\includegraphics[width=8.4cm,angle=0,clip=true]{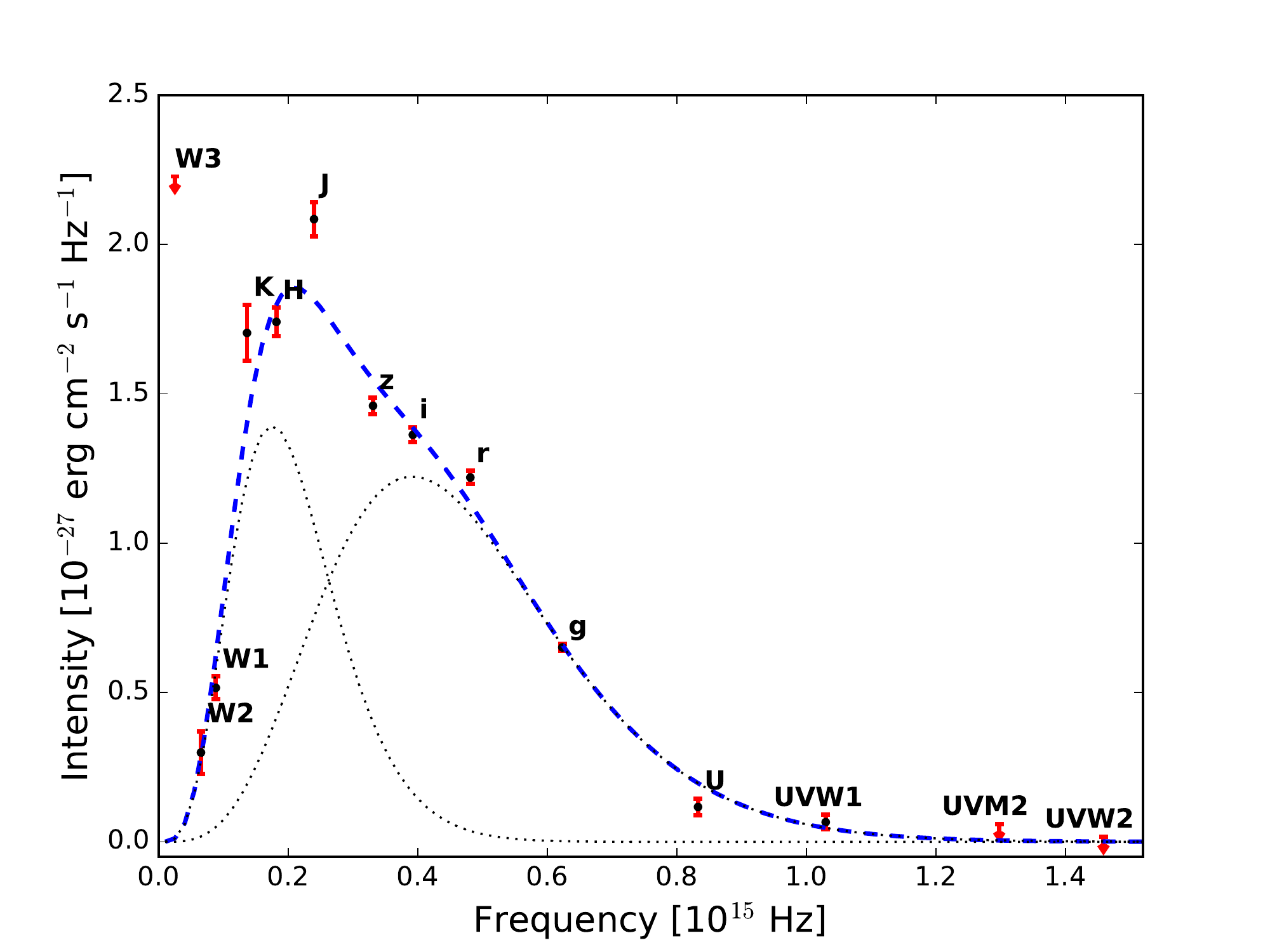}\hspace{-.5cm}}
\caption{
Top to bottom: {\em XMM-Newton} light curve of the 3FGL\, J2039.6$-$5618 counterpart, GROND light curve and multi-band spectrum of its optical counterpart from GROND, UVOT, OM, and {\em WISE} data. The X-ray and optical light curves are aligned in phase. The black dotted lines in the bottom panel are the two BB components to the spectrum (blue dashed line) best-fitting the data points (red). Filters are labelled with their names.}
\label{2039}
\end{center}
\end{figure}

Out of the ten LAT sources with optical coverage in the Catalina Sky Survey (Table~\ref{opt}), we could recover the periodicity of the optical counterparts to the two known RB candidates 3FGL\, J0523.3$-$2528 and 3FGL\, J1653.6$-$0158, while for the third one (3FGL\, J2039.6$-$5618) we could not find any clear evidence of periodicity in the Catalina data,  despite finding evidence in the GROND data (see next paragraph). For the other seven LAT sources we could not find either an X-ray source in the 3FGL error box (3FGL\, J1539.2$-$3324 and J1630.2+3733), or we found X-ray sources with no Catalina counterpart (3FGL\, J1744.1$-$7619 and J2112.5$-$3044), or X-ray sources that do have a Catalina counterpart but that is not periodically variable (3FGL\, J1119.9$-$2204 and J1625.1$-$0021). For 3FGL\, J0802.3$-$5610 we found an X-ray source associated with a Catalina counterpart with only a marginal evidence of periodicity that requires to be confirmed by further optical observations. Out of the seven sources that have also (or only) GROND coverage, we could detect a clear periodicity for the optical counterpart to 3FGL\, J2039.6$-$5618 (Salvetti et al.\ 2015). For 3FGL\, J0523.3$-$2528, we could only partially recover the light curve of its optical counterpart, obtained from the Catalina data, owing to a sparse phase coverage of the GROND data. Like 3FGL\, J1539.2$-$3324 (see above), also 3FGL\, J0744.1$-$2523 has no candidate X-ray counterpart, however, despite this, we found a candidate GROND counterpart in the 3FGL error ellipse with  clearly periodic modulation. Both 3FGL\, J1119.9$-$2204 and J1625.1$-$0021 have candidate X-ray counterparts with potential GROND counterparts but they are not periodically variable. Finally, 3FGL\, J1035.7$-$6720 has candidate X-ray counterparts but it has no potential GROND counterparts.

One of the sources in our sample has been identified as a binary MSP (3FGL\, J1630.2$+$3733; Sanpa-Arsa et al., in preparation). In addition, while this work was being finalised, another two (3FGL\, J1035.7$-$6720 and 3FGL\, J1744.1$-$7619) were detected as pulsars by Clark et al.\ (2017), although no value of the spin (and orbital) period has been published, leaving unspecified whether they are MSPs or not and whether they are binary or isolated. None of the sources in our sample has a radio association in the recent follow-up by Schinzel et al.\ (2017). In the following, we describe the results on a case by case basis.

\subsection{High-confidence Binary MSP candidates}\label{sec:3.1}

\subsubsection{3FGL\, J0523.3$-$2528}

An interesting X-ray candidate counterpart to this originally unassociated $\gamma$-ray source was found by Acero et al.\ (2013) in a 4.8 ks \sw\ observation (see also Takeuchi et al.\ 2013). This is the only X-ray observation available for this source. Soon after, based on an analysis of the Catalina V-band data and on follow-up radial velocity spectroscopy observations, Strader et al.\ (2014) found a periodically modulated optical counterpart (0.688 d), with the light curve featuring two peaks. This set the case for the identification of 3FGL\, J0523.3$-$2528 as a new binary MSP candidate, possibly a RB. In the early phases of our project, we independently found the optical periodicity in this source using the same Catalina data set as used in Strader et al.\ (2014). This triggered a proposal for follow-up GROND observations submitted for the September 2014/March 2015 semester. The GROND observations were eventually performed in February 2015. Unfortunately, the non optimal scheduling of the requested 
observation sequence did not allow us to uniformly cover in phase the 0.688 d cycle, with most measurements covering about half of the two peaks in the light curve.

Fitting the X-ray spectrum with a PL model with $\Gamma_X=1.63\pm0.2$ and N$_H$ fixed to the integrated Galactic value of $1.9\times10^{20}$ cm$^{-2}$, we obtain an unabsorbed 0.3--10 keV flux $F_X=(1.85\pm0.3)\times10^{-13}$ erg cm$^{-2}$ s$^{-1}$. For a $\gamma$-ray flux above 100 MeV of $F_{\gamma}=(1.99\pm0.12)\times10^{-11}$ erg cm$^{-2}$ s$^{-1}$ (Acero et al.~2015), the $\gamma$-to-X flux ratio is $F_{\gamma}/F_X\sim100$.

\subsubsection{3FGL\, J1653.6$-$0158}

Candidate X-ray counterparts to 3FGL\, J1653.6$-$0158 were found by Cheung et al.\ (2012) using a \chan-ACIS 21 ks observation (OBSID 11787). An optical follow-up of the 3FGL\, J1653.6$-$0158 field by Romani et al.\ (2014) led to the discovery of a periodic flux modulation (0.052 d) in the optical counterpart to the brightest of the \chan\ sources, making a case for a new binary MSP identified through the detection of an optical periodicity, also in this case a likely RB. The periodicity was also found in the Catalina data, allowing Romani et al.\ (2014) to extend the time baseline for the period determination. We note that the source (MLS\, J165338.1$-$015836) is not included in the Catalina Surveys Periodic Variable Star Catalogue (Drake et al.\ 2014) since this includes data from the CSS only and not from the MLS. The detection of periodicity in the optical counterpart to the X-ray source was confirmed by Kong et al.\ (2014) from the analysis of independent observations. The spectrum of the X-ray source was fitted with a PL with $\Gamma_{\rm X} =1.65^{+0.39}_{-0.34}$ ($N_{\rm H}=1.3^{+1.8}_{-1.3}\times 10^{21}$ cm$^{-2}$) in the 0.5--8 keV energy range (Romani et al.\ 2014). These values yield an unabsorbed 0.3--10 keV X-ray flux $F_{\rm X}=2.3^{+0.9}_{-0.6} \times 10^{-13}$ erg cm$^{-2}$ s$^{-1}$. Its $\gamma$-ray flux above 100 MeV is $F_{\gamma}=(3.37 \pm 0.18) \times 10^{-11}$ erg cm$^{-2}$ s$^{-1}$ (Acero et al.\ 2015), giving an $F_{\gamma}/F_{\rm X} \sim 150$.

More recently, the \chan\ data have been re-analysed by Hui et al.\ (2015), who claimed possible evidence (at the 99.2\% confidence level) of an X-ray modulation in this source, at the same 0.052 d period as observed in the optical. No evident X-ray modulation, however, was found by Romani et al.\ (2014), whereas only a tentative evidence was found by Kong et al.\ (2014). We reanalysed the same \chan\ data to verify the existence of a possible X-ray modulation. However, our analysis of the folded 0.3--10 keV light curve does not show any evidence of deviation from a constant flux at a level above $3\sigma$. Therefore, our conclusions are in line with those of Romani et al.\ (2014) and Kong et al.\ (2014). Like for 3FGL\, J0523.3$-$2528, also for 3FGL\, J1653.6$-$0158 we independently found in the Catalina data the same periodicity as discovered by Romani et al.\ (2014).

\subsubsection{3FGL\, J2039.6$-$5618}\label{sect_J2039}

As a part of this project, we observed the unassociated {\em Fermi}-LAT source 3FGL\, J2039.6$-$5618 with both \xmm\ (OBSID 0720750301) and GROND and we identified it as a likely RB from the discovery of a common periodicity in its X-ray and optical/near-infrared counterpart at a period of 0.227 d (Salvetti et al.\ 2015), which we identified as the orbital period of a tight binary system, where one of the members is a neutron star.

We also detected the optical counterpart to the X-ray source in the Catalina Sky Survey (217 epochs) but the large error bars attached to the single flux measurements did not allow us to confirm the periodicity observed in the GROND data and search for possible long-term evolution of the light curve. The periodicity has been independently discovered by an analysis of our GROND data by Romani (2015), who complemented them with data taken with the SOAR and Dark Energy Survey (DES) telescopes, allowing him to extend the time baseline for the light curve folding and improve on the determination of the period accuracy. The X-ray and optical light curves are shown in Fig.~\ref{2039} (top and middle panel, respectively), aligned in phase using the updated period determined by Romani (2015), $P_{\rm B} = 0.228116 \pm0.000002$ d, and the epoch of quadrature (MJD=56884.9667$\pm$0.0003) determined by Salvetti et al.\ (2015) by fitting the GROND light curve profile with a geometrical model of the tidally distorted neutron star companion (see also Mignani et al.\, 2016).

In Salvetti et al.\ (2015), we used the GROND data to characterise the phase-averaged spectrum of the X-ray source counterpart together with optical/ultraviolet data from the {\em XMM-Newton} Optical Monitor (OM; Mason et al.\ 2001) and the {\em Swift} Ultraviolet Optical Telescope (UVOT; Roming et al.\ 2005). To these data, we added mid-infrared flux measurements from the archival {\em Wide-Field Infrared Survey Explorer} ({\em WISE}; Wright et al.\ 2010) data, obtained in the W1 (3.4$\mu$m), W2 (4.6$\mu$m), W3 (12$\mu$m), W4 (22$\mu$m) bands, and available in the AllWISE catalogue (Wright et al.\ 2010). The multi-band UV--to--mid-IR spectrum is shown in Fig.~\ref{2039} (bottom).
Interestingly, as qualitatively shown in Mignani et al.\ (2016), the new {\em WISE} data confirm that the source spectrum features a cold black body (BB) with effective temperature $T_{\rm eff}=1700 \pm 120$ K, dominating in the near/mid-infrared, and a hot BB, with $T_{\rm eff}= 3800 \pm 150$ K, dominating in the optical/ultraviolet. We looked for a periodic flux modulation in the multi-epoch {\em WISE} data but we found no evidence of it, within the statistical uncertainty of the {\em WISE} flux measurements ($\sim 0.2$--0.5 magnitudes in the W1 band). As we anticipated in Mignani et al.\ (2016), the lack of a modulation at the optical/near-IR period would suggest that the mid--IR emission component of the spectrum is not produced by the tidally-distorted companion star but by a different source in the binary system.

The spectrum of the 3FGL\, J2039.6$-$5618 X-ray counterpart was fitted with a PL ($\Gamma_{\rm X} =1.36 \pm 0.09$; $N_{\rm H} < 4 \times 10^{20}$ cm$^{-2}$), with an unabsorbed 0.3--10 keV X-ray flux $F_{\rm X}=10.19^{+0.87}_{-0.82} \times 10^{-14}$ erg cm$^{-2}$ s$^{-1}$ (Salvetti et al.\ 2015), which gives $F_{\gamma}/F_{\rm X} \sim 170$ for $F_{\gamma}=(1.71 \pm 0.14) \times 10^{-11}$ erg cm$^{-2}$ s$^{-1}$ (Acero et al.\ 2015).

\onecolumn
\begin{figure*}
\begin{center}
{\includegraphics[width=6.8cm,angle=0,clip=true]{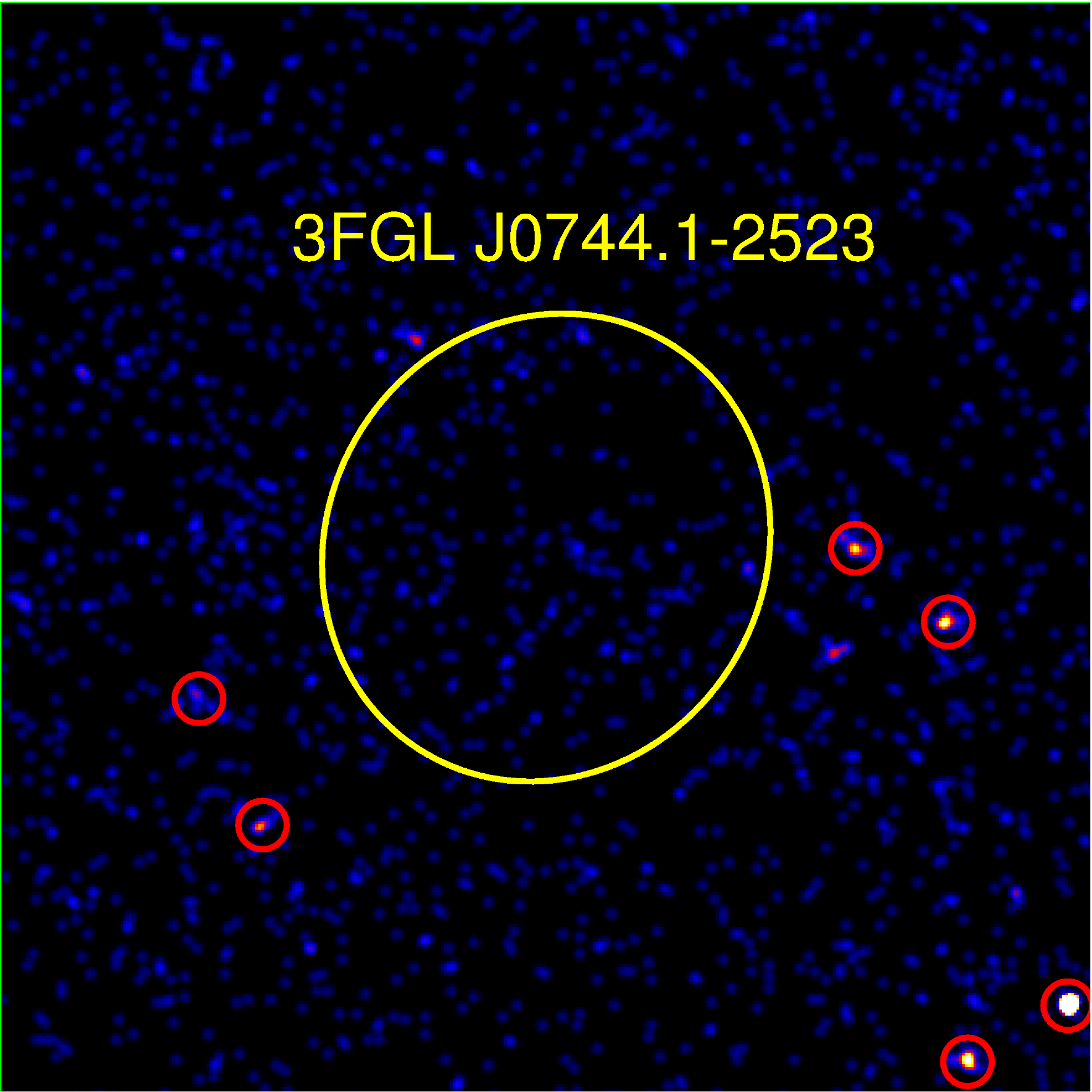}\hspace{.5cm}
\includegraphics[width=6.8cm, angle=0,clip=true]{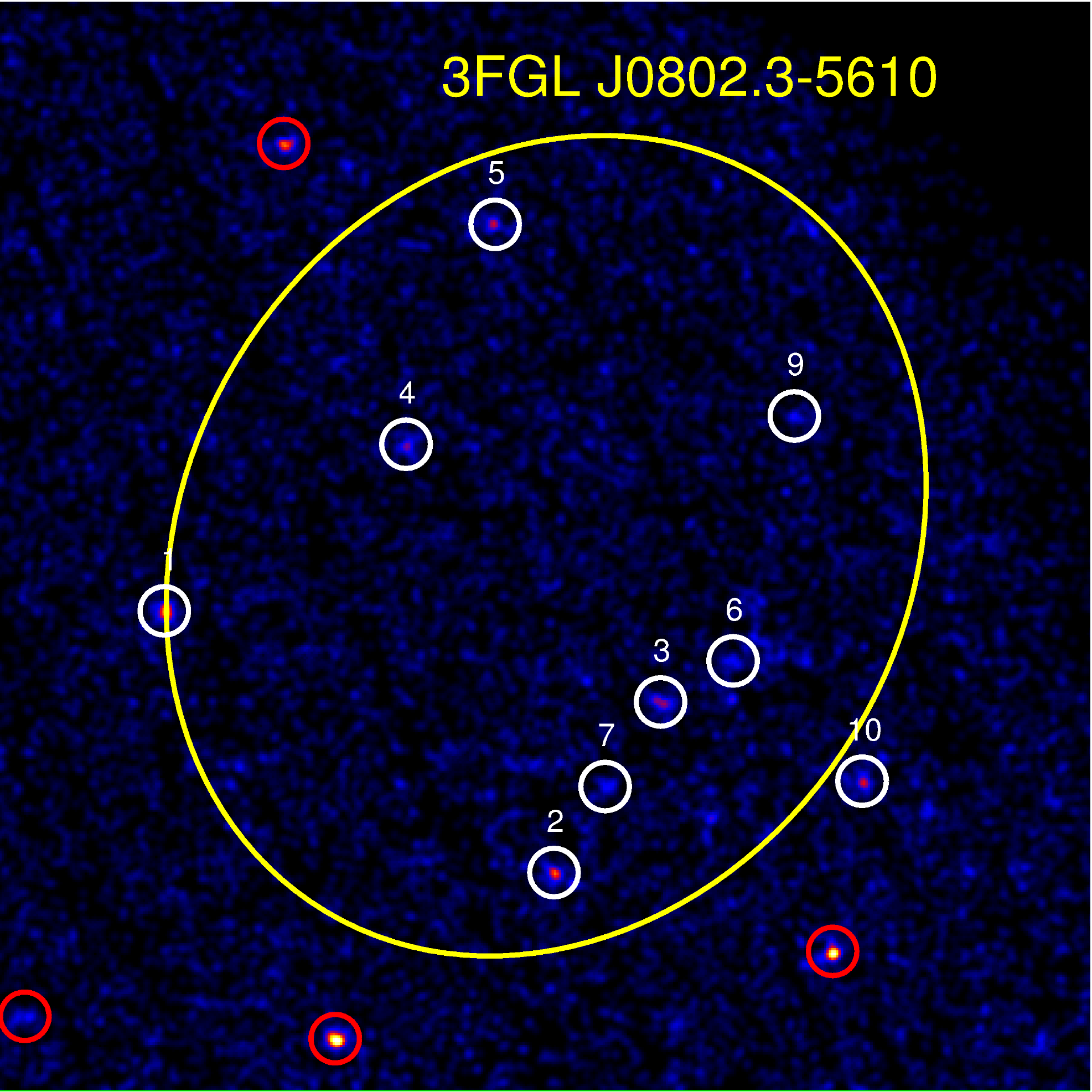}}\\
\vspace{.5cm}
{\includegraphics[width=6.8cm,angle=0,clip=true]{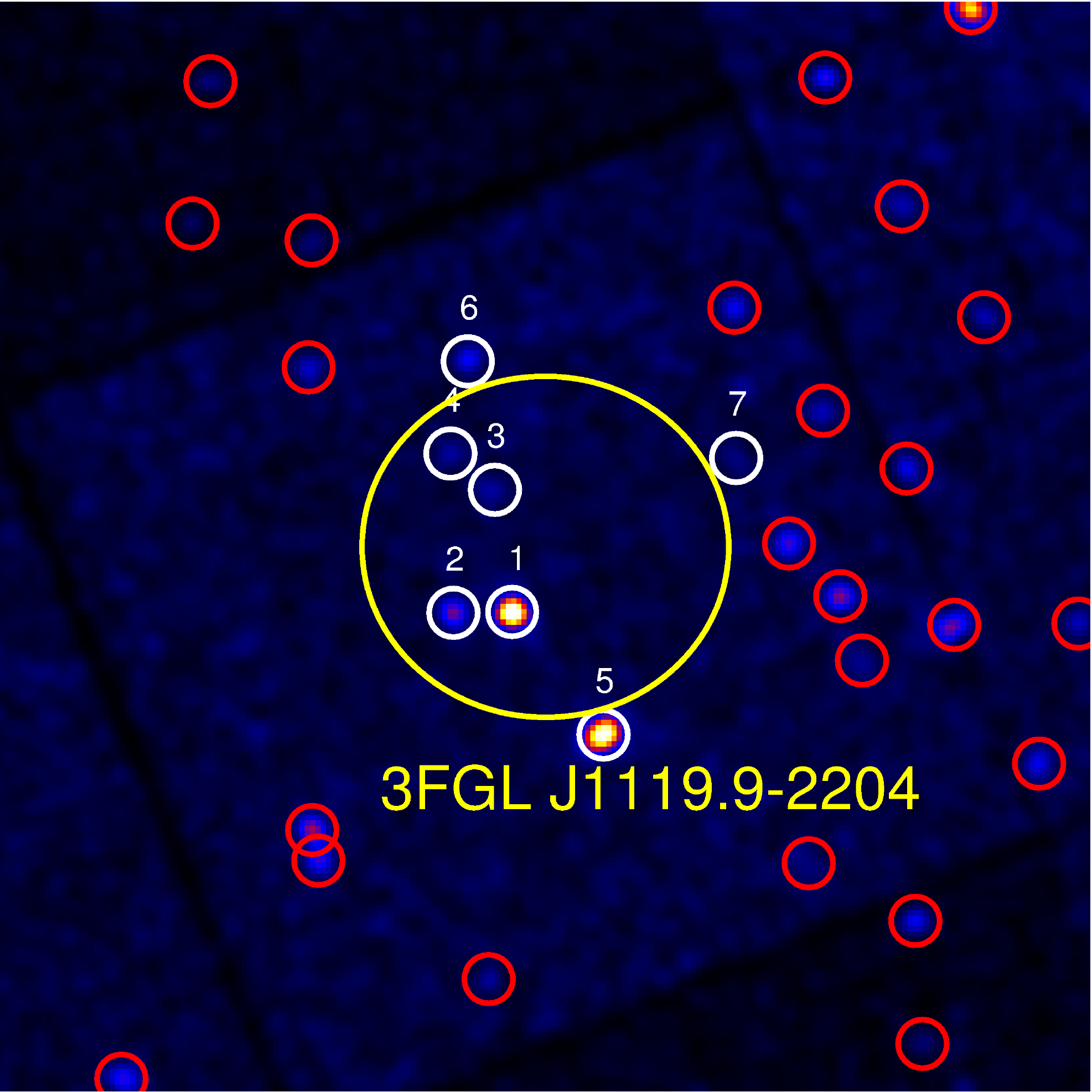}\hspace{.5cm}
\includegraphics[width=6.8cm,angle=0,clip=true]{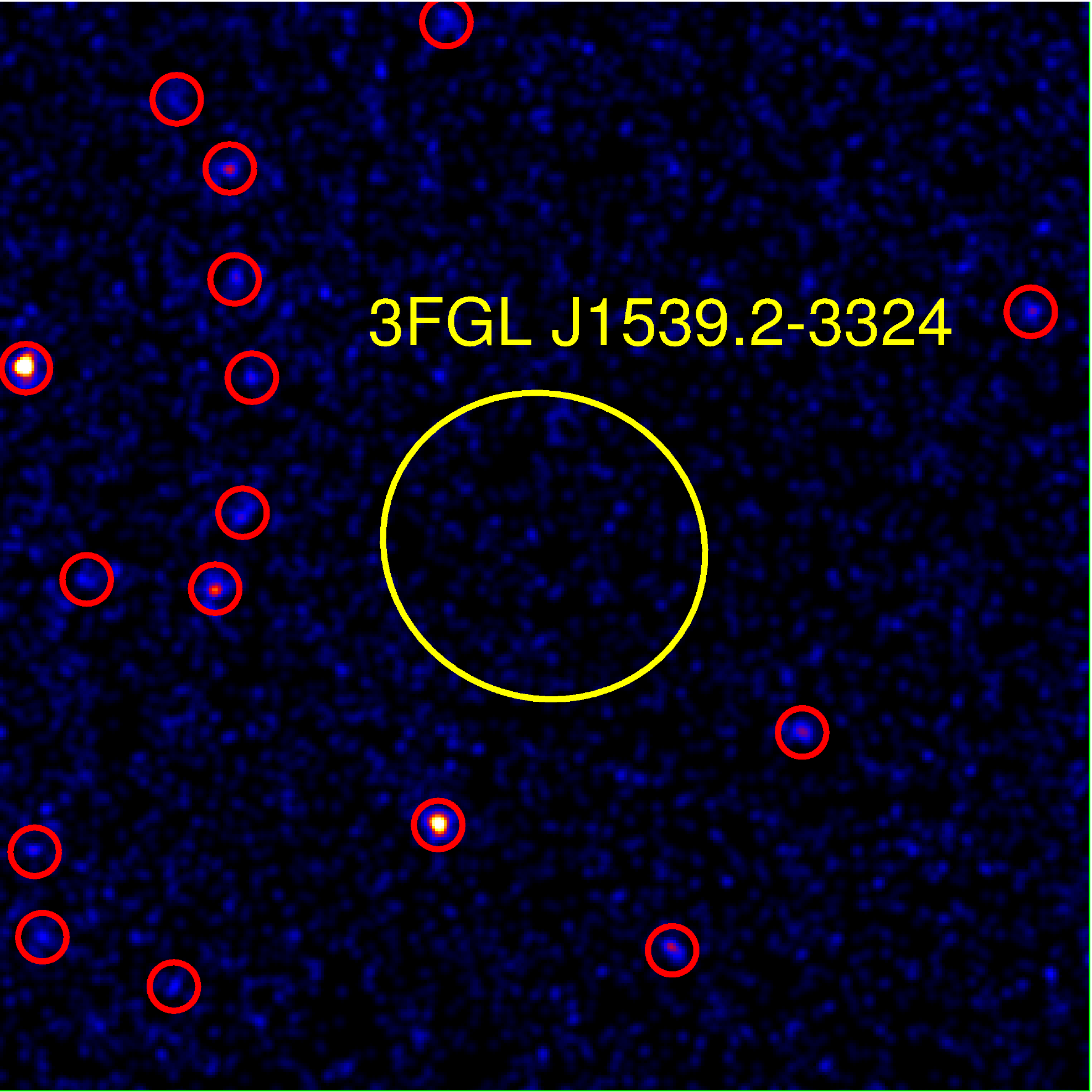}}\\
\vspace{.5cm}
{ \hspace{.1cm}\includegraphics[width=6.8cm,angle=0,clip=true]{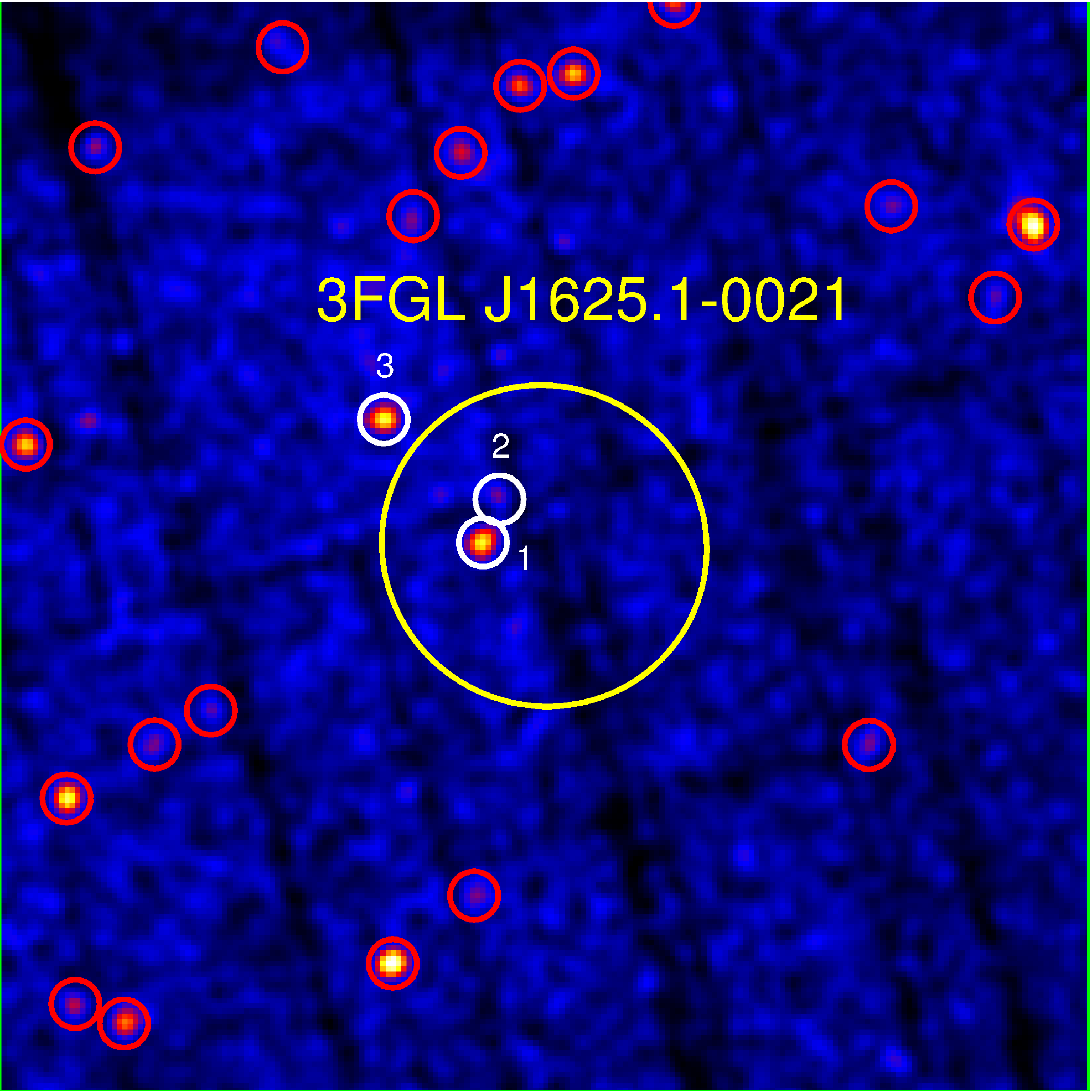}\hspace{.5cm}
\includegraphics[width=6.8cm,angle=0,clip=true]{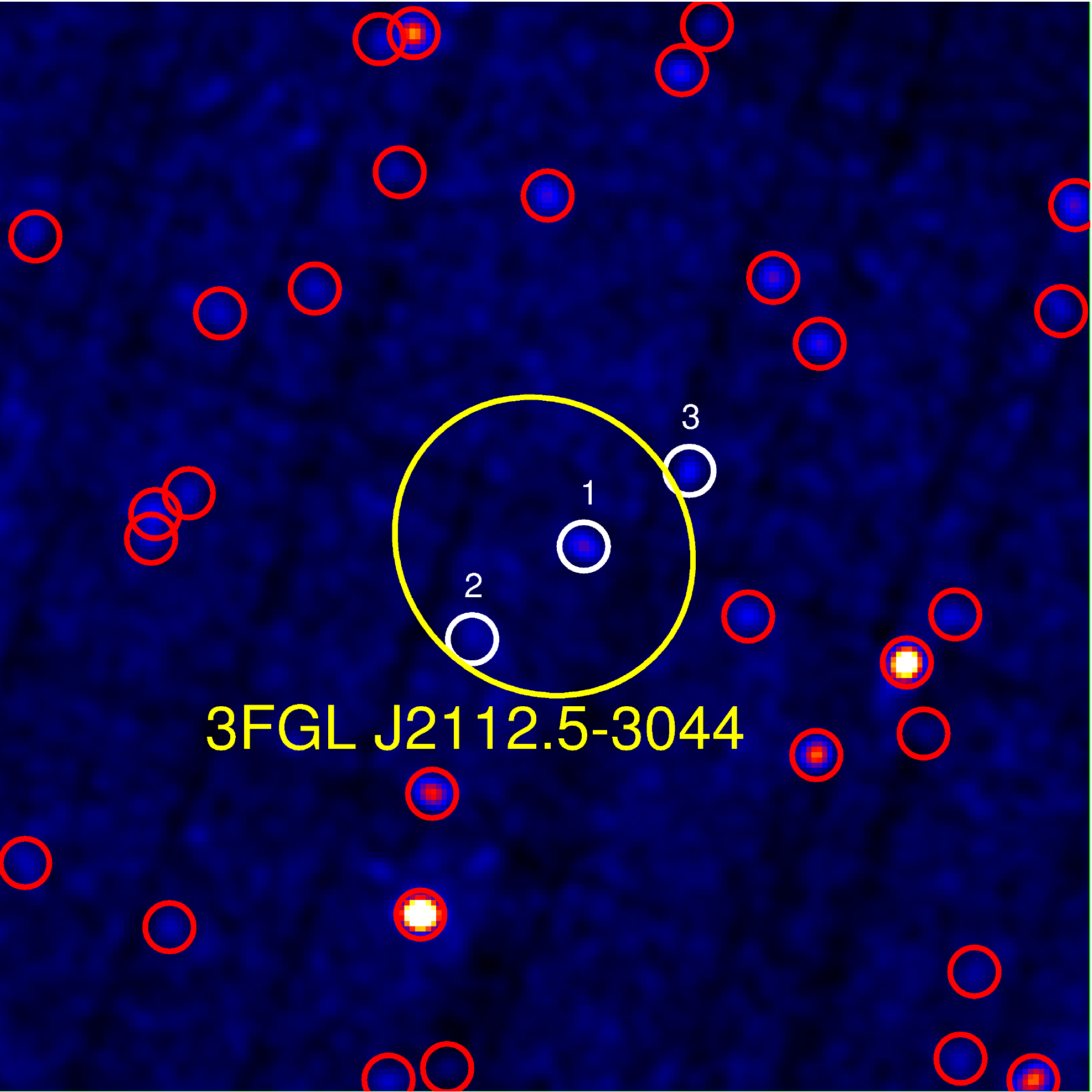}}
\caption{0.3--10 keV exposure-corrected X-ray images ($15\arcmin \times15\arcmin$) of the binary MSP candidates discussed in Sectn.\, 3.2 obtained with the \xmm\ EPIC and \sw\ XRT (3FGL\, J0744.1$-$2523 and 3FGL\, J1539.2$-$3324) instruments. All images have been smoothed using a Gaussian filter with a kernel radius of 3\arcsec. Images include the 95\% confidence level \emph{Fermi}-LAT position error ellipse (shown in yellow), as given in the 3FGL catalogue (Acero et al.\ 2015). Plausible X-ray counterparts to the $\gamma$-ray source, detected within or close to the LAT error ellipse, are highlighted with a white circle and labeled as in Table~\ref{tab2}, while other X-ray sources detected in the FoV are plotted with a red circle.}
\label{XrayImages}
\end{center}
\end{figure*}

\begin{landscape}
\begin{longtable}{lcccccccc}
\caption{Summary of the X-ray parameters of the sources detected within/close to the error ellipse of the {\em Fermi}-LAT source (see Fig. \ref{XrayImages}), as discussed in the text.\\ Here, we report the name of the 3FGL unassociated source, and for each X-ray plausible counterpart the best-fit position, the best-fit column density and photon index, the unabsorbed X-ray flux in the 0.3--10 keV energy band, and the $\gamma$--to--X-ray flux ratio. All uncertainties are reported at the 68\% confidence level. For each observation, we also report the $3 \sigma$ detection limits. In the last column, we flag the association with an optical counterpart to the X-ray source, detected by either Catalina or GROND, and flag whether it features a periodical modulation. \label{tab2}}\\
\hline \hline
\vspace{-3mm}\\
\multirow{2}{*}{3FGL Name} & X-ray & J2000 X-ray coord. & N$_\textrm{H}$ & $\Gamma_{X}$ & Flux$_{(\rm 0.3-10\,keV)}$ & \multirow{2}{*}{$\displaystyle\frac{\mbox{F$_{\gamma}$}}{\mbox{F$_X$}}$} & Detection Limit & Optical\\
 & Source & RA, Dec [$^{\circ}$] (stat. err.$^a$) & $10^{22}$ cm$^{-2}$ &  & $10^{-14}$\,erg cm$^{-2}$ s$^{-1}$ & & $10^{-14}$\,erg cm$^{-2}$ s$^{-1}$ & Counterparts\\
\vspace{-3mm}\\
\hline
\vspace{-3mm}\\
J0744.1$-$2523 & -- & -- & -- & -- & -- & $>$525 & $4.54$ & Yes$^{\rm P}$ \\
\vspace{-3mm}\\
\hline
\vspace{-3mm}\\
\multirow{10}{*}{J0802.3$-$5610} & 1 & 120.7396, --56.1838 (0\farcs92) & \multirow{10}{*}{0.18$^b$} & 1.82$^{+0.32}_{-0.30}$ & 4.93$^{+1.46}_{-1.16}$ & 264$^{+82}_{-67}$ & \multirow{10}{*}{$0.96^c$} & No\\
 & 2 & 120.5795, $-$56.2436 (1\farcs15) &  & 2.17$^{+0.48}_{-0.44}$ & 2.60$^{+1.20}_{-0.80}$ & 500$^{+235}_{-161}$ &  & No\\
 & 3 & 120.5358, $-$56.2045 (1\farcs65) &  & 2.08$^{+0.61}_{-0.51}$ & 2.66$^{+1.40}_{-0.77}$ & 489$^{+261}_{-148}$ &  & No\\
 & 4 & 120.6404, $-$56.1456 (1\farcs56) &  & 1.39$^{+0.54}_{-0.49}$ & 3.10$^{+1.59}_{-1.31}$ & 419$^{+219}_{-181}$ &  & No\\
 & 5 & 120.6040, $-$56.0952 (1\farcs20) & & 4.08$^{+0.74}_{-0.68}$ & 9.62$^{+5.72}_{-3.63}$ & 135$^{+81}_{-52}$ &  & Yes$^{\rm P}$ \\
 & 6 & 120.5061, $-$56.1951 (1\farcs59) &  & 2$^b$ & 2.18$^{+1.99}_{-1.74}$ & 596$^{+547}_{-479}$ &  & No\\
 & 7 & 120.5585, $-$56.2239 (1\farcs98) &  & 2$^b$ & 1.76$^{+1.56}_{-1.35}$ & 739$^{+658}_{-571}$ &  & No\\
 & 8 & 120.4751, $-$56.0959 (1\farcs98) &  & 2$^b$ & 1.78$^{+1.27}_{-1.12}$ & 730$^{+525}_{-464}$ &  & No\\
 & 9 & 120.4813, $-$56.1389 (1\farcs95) &  & 2$^b$ & 1.67$^{+1.45}_{-1.28}$ & 778$^{+680}_{-601}$ &  & No\\
 & 10 & 120.4529, $-$56.2225 (1\farcs41) &  & 2.40$^{+0.50}_{-0.45}$ & 2.50$^{+1.34}_{-0.75}$ & 520$^{+283}_{-163}$ &  & No\\
\vspace{-3mm}\\
\hline
\vspace{-3mm}\\
\multirow{7}{*}{J1119.9$-$2204} & 1 & 169.9927, $-$22.0822 (0\farcs38) & \multirow{7}{*}{0.04$^b$} & 2.63$^{+0.12}_{-0.11}$ & 7.29$^{+0.56}_{-0.54}$ & 230$^{+22}_{-22}$ & \multirow{7}{*}{$0.41^c$} & Yes \\
 & 2 & 170.0072, $-$22.0824 (0\farcs72) &  & 1.99$^{+0.18}_{-0.17}$ & 1.97$^{+0.35}_{-0.32}$ & 853$^{+160}_{-148}$ &  & Yes$\dagger$ \\
 & 3 & 169.9970, $-$22.0543 (1\farcs79) &  & 1.70$^{+0.35}_{-0.33}$ & 1.44$^{+0.49}_{-0.37}$ & 1167$^{+403}_{-308}$ &  & No\\
 & 4 & 170.0079, $-$22.0460 (1\farcs43) &  & 2$^b$ & 0.43$^{+0.21}_{-0.20}$ & 3907$^{+1922}_{-1832}$ &  & No\\
 & 5 & 169.9702, $-$22.1102 (0\farcs40) &  & 2.41$^{+0.09}_{-0.09}$ & 8.05$^{+0.65}_{-0.57}$ & 209$^{+21}_{-19}$ & &  No\\
 & 6 & 170.0036, $-$22.02489 (\farcs05) &  & 1.69$^{+0.26}_{-0.25}$ & 1.43$^{+0.38}_{-0.32}$ & 1175$^{+320}_{-272}$ &  & Yes$\ddagger$ \\
 & 7 & 169.9375, $-$22.0470 (1\farcs92) &  & 2$^b$ & 0.58$^{+0.19}_{-0.22}$ & 2897$^{+964}_{-1112}$ & & No\\
\vspace{-3mm}\\
\hline
\vspace{-3mm}\\
J1539.2$-$3324 & -- & -- & -- & -- & -- & $>$1350 & 0.85 & --\\
\vspace{-3mm}\\
\hline
\vspace{-3mm}\\
\multirow{3}{*}{J1625.1$-$0021} & 1 & 246.2935, --0.3578 (0\farcs91) & \multirow{3}{*}{0.07$^b$} & 3.19$^{+0.26}_{-0.25}$ & 2.22$^{+0.40}_{-0.34}$ & 824$^{+158}_{-137}$ & \multirow{3}{*}{$0.62^c$} & No\\
 & 2 & 246.3162, $-$0.3296 (1\farcs75) & & 2$^b$ & 0.64$^{+0.40}_{-0.39}$ & 2859$^{+1797}_{-1752}$ &  & Yes$\ddagger$ \\
 & 3 & 246.2897, $-$0.3480 (0\farcs85) &  & 1.53$^{+0.21}_{-0.20}$ & 2.94$^{+0.73}_{-0.57}$ & 622$^{+160}_{-127}$ &  & Yes \\
\vspace{-3mm}\\
\hline
\vspace{-3mm}\\
\multirow{3}{*}{J2112.5$-$3044} & 1 & 318.1341, --30.7344 (1\farcs03) & \multirow{3}{*}{0.07$^b$} & 2.59$^{+0.29}_{-0.28}$ & 1.41$^{+0.39}_{-0.22}$ & 1348$^{+386}_{-233}$ & \multirow{3}{*}{$0.36^c$} & No\\
& 2 & 318.1060, $-$30.7171 (2\farcs48) &  & 1.91$^{+0.52}_{-0.51}$ & 0.66$^{+0.49}_{-0.24}$ & 2879$^{+2148}_{-1068}$ &  & No\\
& 3 & 318.1638, $-$30.7556 (1\farcs07) &  & 2.16$^{+0.51}_{-0.45}$ & 0.79$^{+0.45}_{-0.21}$ & 2405$^{+1381}_{-663}$ &  & No\\
\vspace{-3mm}\\
\hline\\
\multicolumn{9}{l}{\footnotesize{$^a$Here we report only the $1\sigma$ statistical error, the $1\sigma$ systematic uncertainty is 1\farcs5 for X-ray sources detected by {\em XMM-Newton}.}}\\
\multicolumn{9}{l}{\footnotesize{$^b$Owing to the low statistics in these sources, we fixed this parameter in the spectral analysis.}}\\
\multicolumn{9}{l}{\footnotesize{$^c$Here we computed the $3\sigma$ detection limit combining data from the pn, MOS1 and MOS2 detectors, as described in Baldi et al.\ (2002).}}\\
\multicolumn{9}{l}{\footnotesize{$^\dagger$Identified as a galaxy}}\\
\multicolumn{9}{l}{\footnotesize{$^\ddagger$Two possible counterparts}}
\end{longtable}
\end{landscape}

\twocolumn
\subsection{Possible binary MSP candidates}\label{sec:3.2}

\subsubsection{3FGL\, J0744.1$-$2523}

\begin{figure}
\begin{center}
{\hspace{-0.6cm}\includegraphics[width=7.8cm,angle=0,clip=true]{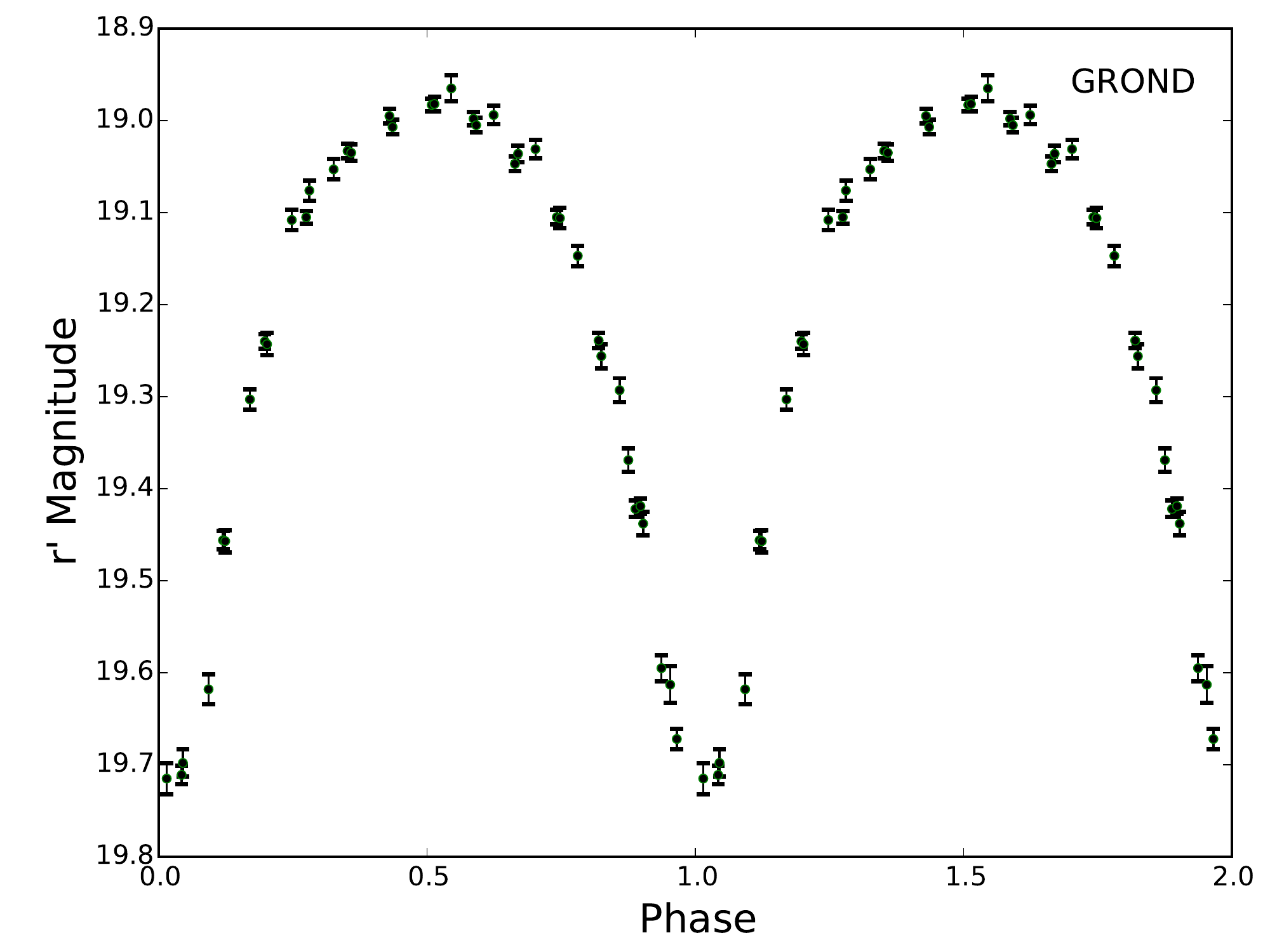}}
{\includegraphics[width=8.4cm,angle=0,clip=true]{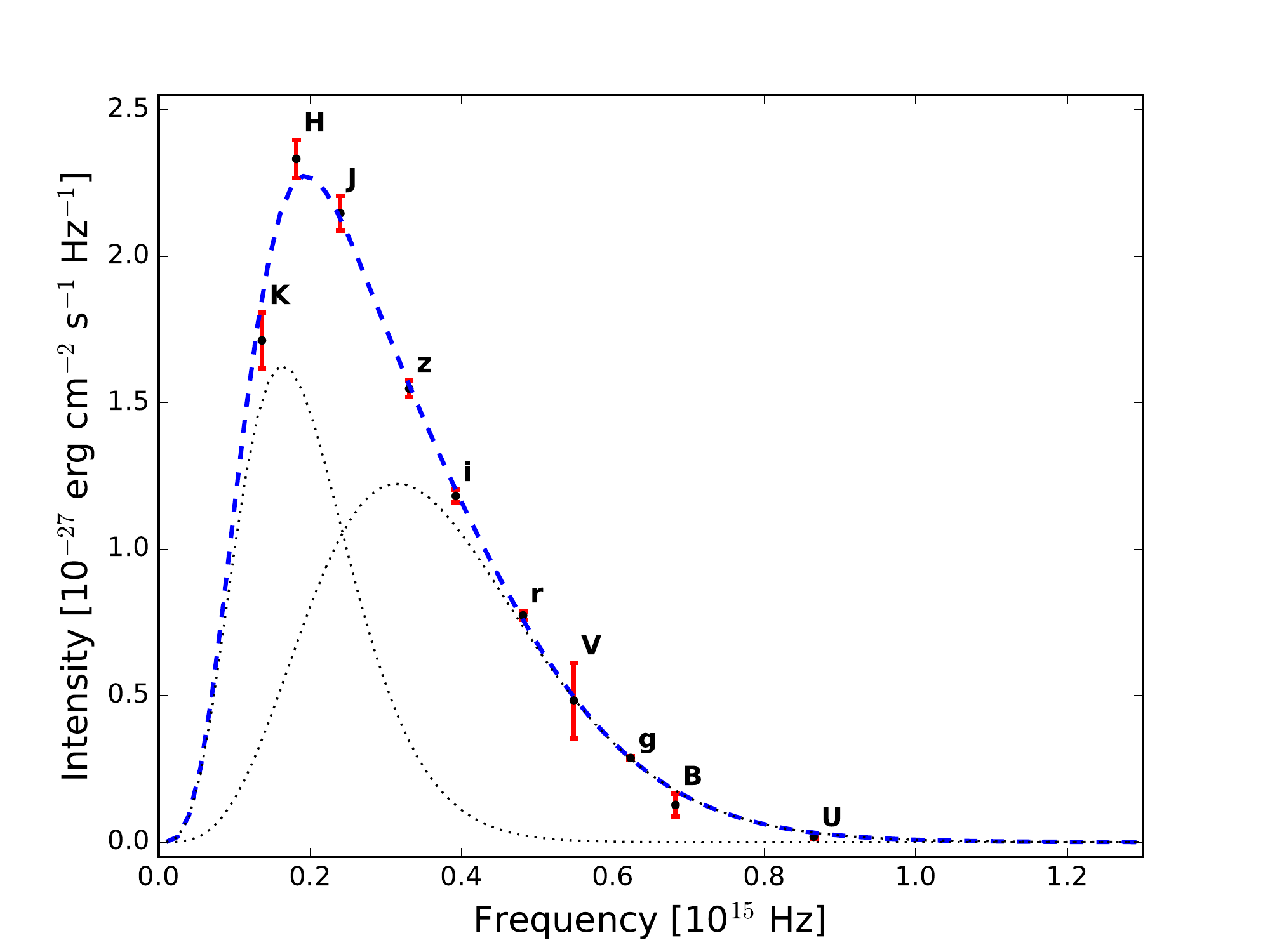}\hspace{.5cm}}
\caption{
Top to bottom: GROND light curve and spectrum of the 3FGL\, J0744.1$-$2523 candidate counterpart. In the bottom panel, the two BB components to the best-fit spectrum (blue dashed line) are shown as black dotted lines. Observed fluxes are shown by the red points. Filters are labelled with their names (updated from Mignani et al.\ 2016). No interstellar extinction correction has been applied. All magnitudes are in the AB system.}
\label{0744}
\end{center}
\end{figure}

The field of 3FGL\, J0744.1$-$2523 is not covered by the Catalina Sky Survey. However, we found an interesting GROND source within the 3FGL error ellipse (time-averaged magnitude $r'\sim19.18$), at $\alpha=07^h44^m08\fs47$ and $\delta=-25^{\circ}23\arcmin 58\farcs9$ (J2000), which is clearly variable (H\"usemann 2015). This source features a clear flux modulation with an optical period equal to 0.11542$\pm$0.00005 d and an amplitude of $\sim 0.8$ magnitudes (Fig.~\ref{0744}, top). We estimated the period uncertainty following Gilliland et al. (1987). The significance of this modulation is 7.4$\sigma$, as computed from the Generalised Lomb-Scargle periodogram algorithm (Zechmeister \& Kuerted 2009). We recognised the same modulation in the near-IR bands, where the best period is found at 0.11546$\pm$0.00009 d. Therefore, this source could be a candidate companion to a MSP in a tight binary system.

We did not detect any associated X-ray source in the {\em Swift} observations (23 ks total integration), down to a 0.3--10 keV flux limit $F_{\rm X} = 4.5 \times 10^{-14}$ erg cm$^{-2}$ s$^{-1}$ ($3 \sigma$ level). For a $\gamma$-ray flux $F_{\gamma} = (2.38 \pm 0.17) \times 10^{-11}$ erg cm$^{-2}$ s$^{-1}$ (Acero et al.\ 2015), this would correspond to a $\gamma$-to-X-ray flux ratio $F_{\gamma}/F_{\rm X} \ga 525$ for 3FGL\, J0744.1$-$2523. No other X-ray observations of this field have ever been performed. 

\begin{figure*}
{\includegraphics[width=7cm,angle=0,clip=true]{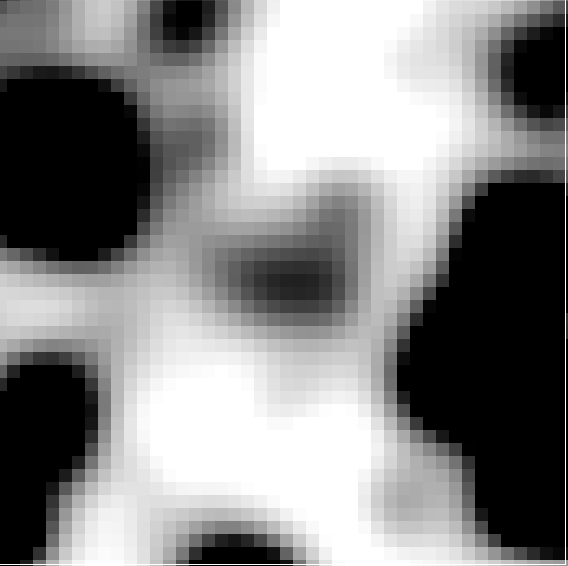}}\hspace{1cm}
{\includegraphics[width=7cm,angle=0,clip=true]{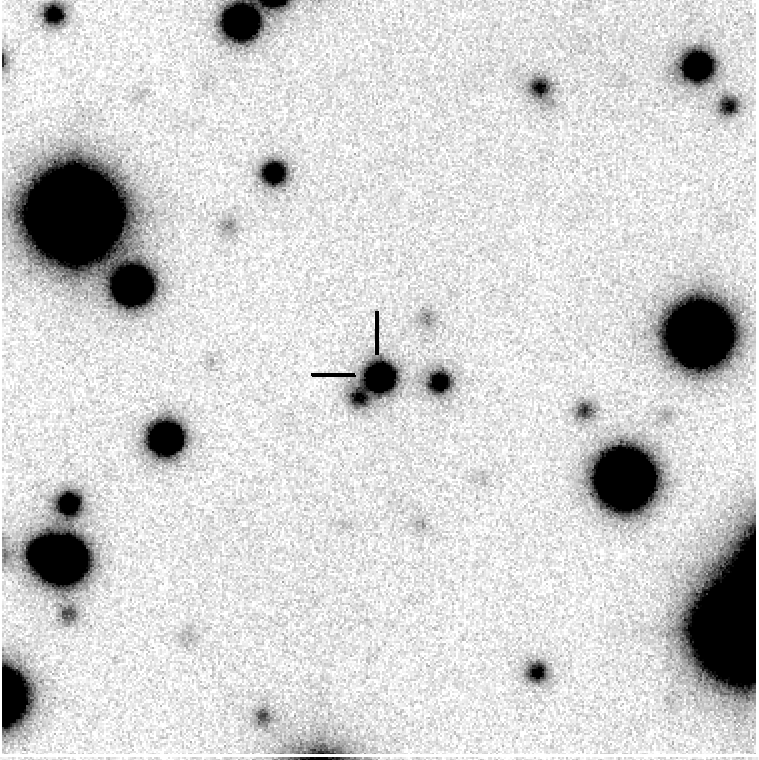}}
\caption{{\em WISE} W1-band image ($60\arcsec\times60\arcsec$) of the candidate optical counterpart to 3FGL\, J0744.1$-$2523 (left) compared with the GROND g$'$-band image (right). The candidate optical counterpart is marked by the two ticks. The mid-infrared source matched in the {\em AllWISE} catalogue is at the centre of the image and is actually resolved in three or more stars in the GROND image, which has a finer pixel scale (0\farcs158 against 2\farcs75) but covers the same area.}
\label{0744_fc}
\end{figure*}

Interestingly, the spectrum of the GROND source (Fig.~\ref{0744}, bottom) is very similar to that of the optical counterpart to 3FGL\, J2039.6$-$5618 (Fig. \ref{2039}, bottom), with cold and hot BBs at comparable temperatures, possibly suggesting similar characteristics for the companion star, as suggested in Mignani et al.\ (2016). Like we did for 3FGL\, J2039.6$-$5618, to investigate this possible similarity we searched for a mid-infrared counterpart in the archival {\em WISE} data and found that the fluxes reported in the AllWISE catalogue (Wright et al.\ 2010) are well above the composite spectrum best-fitting the GROND data. However, inspection of the {\em WISE} images shows that the mid-infrared source (Fig.\ref{0744_fc}, left) is likely a blending of three (or more) sources which are clearly resolved in the higher spatial resolution GROND data (Fig.\ref{0744_fc}, right). Therefore, the fluxes of the mid-infrared source matched in the {\em WISE} data cannot be used to constrain the spectrum of the GROND source.

To complement the data presented in Mignani et al.\ (2016) and better define the source spectrum at shorter wavelengths, we looked for photometry data in the \sw-UVOT observations. There are four sets of observations taken in the V, B and U bands, the first (obsid 00041337001) taken in August 2010, had only U-band data. Three more observations (OBSIDs 00031960001, 2 and 3) were taken in April 2011 with all the three optical filters. We used the standard \texttt{uvotmaghist} and \texttt{uvotsource} FTOOLs with a 3\arcsec\ aperture at the source position, as computed from the GROND images, and an annulus between 11\farcs6 and 16\farcs8 from the source position for the background (smaller than the standard annulus because of the crowded field), to measure the flux of the source. This was clearly detected in a few hundred seconds of exposure time in both the V and B bands. We could not find any evidence of significant variability across single exposures, so that we decided to co-add all of them to achieve higher signal--to--noise detections, with a corresponding integration time of 3527 s and 9032 s in the B and V bands, respectively. In The U band, we could not detect the source in the single exposures. However, the source is detected in the exposure co-addition (10296 s integration time), although at a marginal signal--to--noise $\sim 3 \sigma$. The results of our photometry are U=23.12$\pm$0.32, B=21.07$\pm$0.16, V=19.85$\pm$0.09, where all magnitudes have been converted into the AB system to be directly comparable with the GROND ones.

\begin{figure}
{\includegraphics[width=8.2cm,angle=0,clip=true]{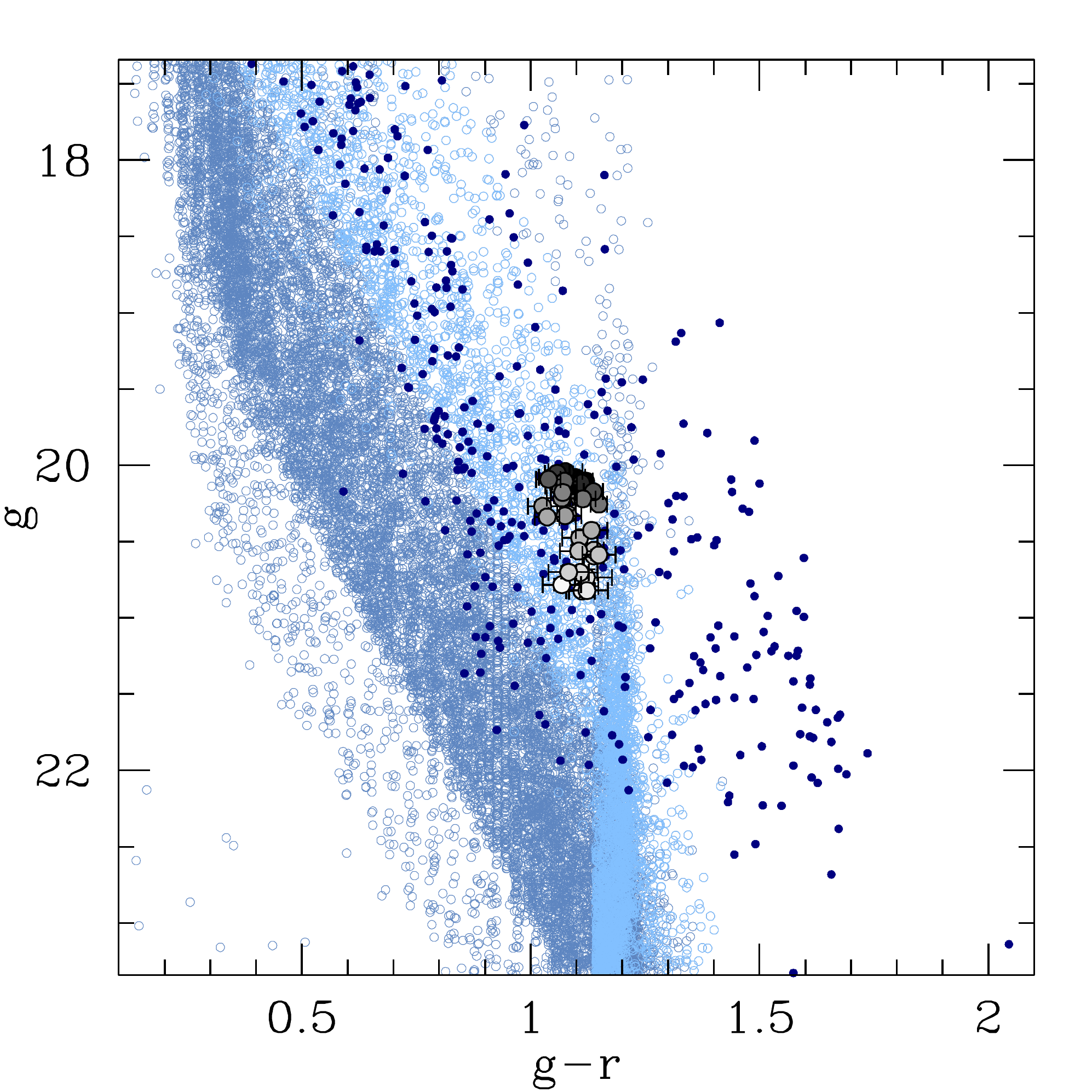}}
\caption{Observed colour-magnitude diagram for the stars detected by GROND in the 3FGL\, J0744.1$-$2523 field (dark blue points). The grey points correspond to the variable GROND source, with the intensity scale varying from the minimum (light grey) to the maximum of the light curve (dark grey). The dark and light cyan points correspond to a simulated stellar population from the Besan\c{c}on models (Robin et al.\ 2004) for different values of the distance, i.e $<$10 kpc and 1--3 kpc, respectively.}
\label{0744_cmd}
\end{figure}

We used the multi-band flux information on the GROND source to determine its nature from its colours. Fig.\ref{0744_cmd} shows the source position in the $g'$ vs $g'-r'$ colour magnitude diagram (CMD), compared to that of the field sources (blue points) and stars simulated with the Besan\c{c}on models (Robin et al.\ 2004) for different distance values, i.e. 1--3 kpc (light cyan) and $<$10 kpc (dark cyan). The grey points correspond to the variable GROND source, with the grey scale varying with the phase of the light curve, i.e. light and dark grey correspond to the phases of the light curve minimum and maximum, respectively. The CMD clearly shows that the position in the diagram fluctuates as a function of the colour evolution along the light curve but remains along the sequence of field stars and is consistent with a late MS star at a distance of 1 to 3 kpc. Note that we neglected the effect of the extinction in the CMD. Such an assumption is justified by the properties of the colour-colour diagram for stars in the GROND field of view. In fact, we compared the observed colour-colour diagram (distance independent) with the expected colour-colour diagram based on the simulations with the Besan\c{c}on models and we found a good agreement between the two without introducing any extinction value. The Galactic hydrogen column density in the source direction is $N_{\rm H} \sim 6\times10^{21}$ cm$^{-2}$ (Dickey \& Lockman 1990), corresponding to E(B-V)$\sim$1.1 (Predehl \& Schmitt 1995). This value is larger than the modest extinction values that we inferred from the colour-colour diagrams, which independently confirms that the star must be relatively close, as suggested by the comparison with the Besan\c{c}on models. For instance, a simple scaling of the $N_{\rm H}$ to produce an interstellar reddening E(B-V)$\sim$0.11, i.e. one tenth of the integrated Galactic value, would put the source at a distance of $\approx$ 1.5 kpc, perfectly consistent with the distance range inferred from the Besan\c{c}on models. We note that this object appears in the first {\em Gaia} data release\footnote{{\tt http://www.cosmos.esa.int/web/gaia/dr1}} but, unfortunately, it has no measured parallax or proper motion yet.

\subsubsection{3FGL\, J0802.3$-$5610}\label{sect_J0802}

\begin{figure}
{\includegraphics[width=8.5cm,angle=0,clip=true]{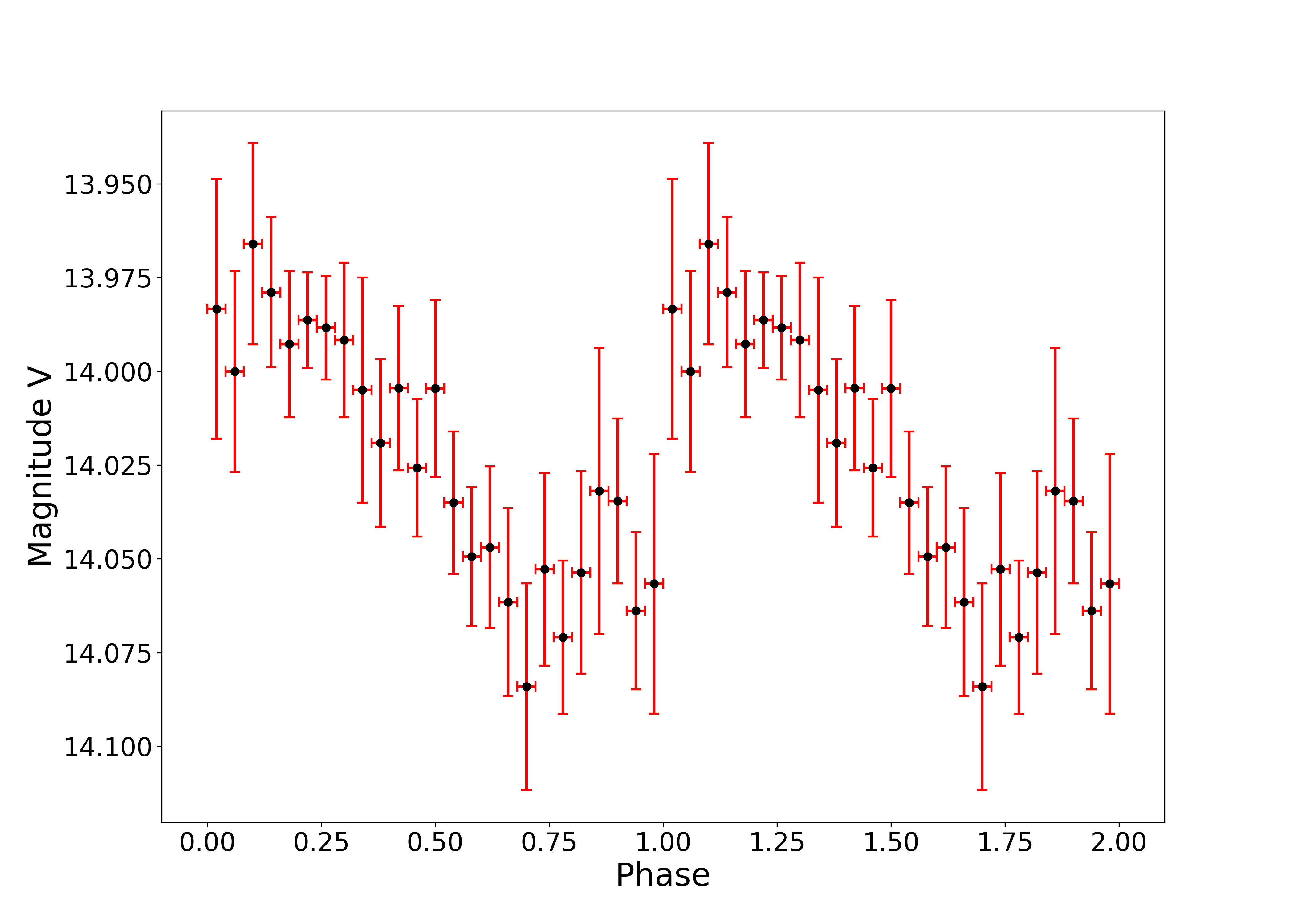}}
\caption{Folded Catalina light curve of the star SSS\, J080225.1$-$560543 that is the optical counterpart of the \xmm\ {\it Source \#5}, possibly associated with 3FGL\, J0802.3$-$5610. The light curve has been folded at the best period of $\sim$0.4159 d. A rebinning of 0.04 in phase has been applied for a better visualisation.}
\label{0802}
\end{figure}

The field of 3FGL\, J0802.3$-$5610 has been observed by both \sw\ (8 ks) and \xmm\ (18 ks). From the \sw\ data we detected no X-ray source at a level of significance above 3$\sigma$ in the 3FGL error ellipse of the $\gamma$-ray source. Therefore, we used the deeper \xmm\ observation (OBSID 0691980301) to search for a candidate X-ray counterpart to 3FGL\, J0802.3$-$5610. For our X-ray analysis, we selected only 0--4 pattern events from the pn and 0--12 events from the two MOS detectors with the default flag {\em mask}. We ran the source detection in the 0.3--10 keV energy range simultaneously on the event lists of each of the EPIC-pn and MOS detectors using a maximum likelihood fitting with the SAS task {\tt edetect\_chain} invoking other SAS tools to produce background, sensitivity, and vignetting-corrected exposure maps. The final source list includes ten X-ray sources close to the 3FGL error ellipse detected from both the pn and MOS detectors, with a combined pn+MOS detection likelihood greater than 10, corresponding to a $\ga$3.5$\sigma$ detection significance.

The \xmm\ image of the 3FGL\, J0802.3$-$5610 field with the detected X-ray sources is shown in Fig.~\ref{XrayImages}. The X-ray source properties are summarised in Table~\ref{tab2}. We detected a Catalina candidate counterpart only for {\it Source \#5}. This is the brightest X-ray source detected within/close to the 3FGL error ellipse, with an unabsorbed flux in the 0.3--10 keV energy range of $F_X=9.6^{+5.7}_{-3.6}\times10^{-14}$ erg cm$^{-2}$ s$^{-1}$. The spectrum is very soft and, using the C-statistics, we could fit it with a PL with photon index $\Gamma_X=4.08^{+0.74}_{-0.68}$ and $N_{\rm H}$ fixed to $1.8\times10^{21}$ cm$^{-2}$. If {\it Source \#5} were the X-ray counterpart to 3FGL\, J0802.3$-$5610, this would have an $F_{\gamma}/F_{\rm X}$ of $\sim135$, for $F_{\gamma} = (1.30 \pm 0.12) \times 10^{-11}$ erg cm$^{-2}$ s$^{-1}$ (Acero et al.\ 2015). Although the PL model provides a statistically acceptable fit (C=5.65 with 2 degrees of freedom), the photon index is unrealistically large. By fitting the spectrum with an absorbed black-body (BB) model with fixed N$_{\rm H}$ we obtained a temperature of $kT=0.15\pm0.03$ keV with C=3.93. The thermal scenario would provide a more reasonable description of the X-ray emission of {\it Source \#5}.

The Catalina counterpart (SSS\, J080225.1$-$560543) to {\it Source \#5} is at an angular distance smaller than $1\arcsec$ ($\alpha=08^h02^m25\fs08$ and $\delta=-56^{\circ}05\arcmin43\farcs8$, J2000) from the best-fit X-ray source position, compatible with the uncertainty on the \xmm\ astrometry. The source is quite bright, with a magnitude V$\sim$14.03 averaged over 218 epochs extending over a time span of about 8 years. By running the Generalised Lomb-Scargle periodogram algorithm on the light curve, we found that the most significant candidate period is at $\sim$1 day with its harmonics. Examining the power spectrum of the window function we verified that these periods are associated with the cadence of the observations. Except for these periods, we detected another distinct peak located at a period of $\sim$0.4159 d, characterised by a significance of $\sim$5.3$\sigma$, after accounting for the number of trials. In addition, we observed several aliases of the candidate frequency at consistent offsets from each harmonic of the daily periodicity. Since the strong daily periodicity affects the statistics in the spectrum, we filtered the time series subtracting the daily sinusoid from the data as suggested by Horne \& Baliunas (1986) to determine whether the tentative period is real. After filtering, we observed that the putative period become the strongest peak in the new periodogram but with a significance just below 5$\sigma$. It is clear that the 1-day signal in the observing window hampers a robust estimate of the statistical significance of the candidate signal. Thus, we cannot draw firm conclusions about the reality of the $\sim$0.4159 d modulation. Fig. \ref{0802} shows the Catalina light curve folded at the period of $\sim$0.4159 d, which seems to feature a single, possibly asymmetric, peak. The field of 3FGL\, J0802.3$-$5610 has not been observed by GROND, so that we cannot confirm the putative periodicity seen in the Catalina data. Because of the poor counts statistic for {\it Source \#5} ($\sim$40 net counts), we could not search for an X-ray flux modulation at the Catalina period (nor any other significant type of variation) in the \xmm\ data.

Owing to the association with a potentially periodic variable optical counterpart, {\it Source \#5} is a promising counterpart to 3FGL\, J0802.3$-$5610. However, the lack of information about the optical spectrum of its candidate counterpart, together with the uncertain value of the best-fit period and its soft, likely thermal, X-ray spectrum prevent us from discarding other plausible scenarios, such as a W UMa or a $\beta$ Lyr binary system or a cataclysmic variable. Deeper X-ray observations and a better spectral and timing characterisation of the optical counterpart are needed to determine the nature of the X-ray source.

Other X-ray sources detected within/close to the $\gamma$-ray error ellipse are fainter than {\it Source \#5} and are characterised by a $\gamma$--to--X-ray flux ratio in the range $\sim$260--780. None of them is either associated with a candidate Catalina counterpart or shows a significant X-ray time variability during the \xmm\ observation. We computed the 3$\sigma$ detection limit by combining data from the pn, MOS1 and MOS2 detectors, as described in Baldi et al.\ (2002), corresponding to an unabsorbed X-ray flux of $9.6\times10^{-15}$ erg cm$^{-2}$ s $^{-1}$.

\subsubsection{3FGL\, J1119.9$-$2204}\label{sect_J1119}

The field of 3FGL\, J1119.9$-$2204 was observed by both \sw\ (67 ks) and \xmm\ (73 ks). The analysis of the \sw\ observations was already reported by Hui et al.\ (2015), while here we focus on the unpublished \xmm\ ones. The \xmm\ observations (OBSID 0742930101) have been performed using the pn detector in fast timing mode to search for pulsations in the brightest X-ray sources detected by \sw, whereas the MOS detectors were used in full-frame mode.

We combined the data from both MOS detectors to increase the sensitivity of the source detection, which we carried out as described in the previous sections. We found more sources close to the 3FGL error ellipse in the \xmm\ MOS1+MOS2 observation with respect to the \sw\ observation (Hui et al.\ 2015). The final source list includes 7 X-ray sources, with a combined detection significance above $3.5\sigma$. Fig.~\ref{XrayImages} shows the 0.3--10 keV exposure-corrected \xmm\ FoV obtained combining the images of both MOS detectors. X-ray source properties are summarised in Table~\ref{tab2}. Three of these sources have a plausible Catalina counterpart, which are {\it Source \#1}, {\it Source \#2} and {\it Source \#6}. None of these sources is associated with a clearly periodic optical counterpart in the Catalina data. {\it Source \#1} is coincident with the \sw\ source J1120\_X1 of Hui et al.\ (2015) and its candidate Catalina counterpart (J111958.3$-$220456; V=15.6) was detected by both the {\it Siding Springs Survey} and the {\it Catalina Sky Survey} in 234 and 76 observations, respectively. By running the Lomb-Scargle algorithm on the Catalina data, Hui et. al.\ (2015) identified a series of peaks in the periodogram that they considered to be caused by the non-uniform distribution of the data points. Our analysis of the Lomb-Scargle periodogram confirms their conclusions. {\it Source \#1} is also associated with a GROND counterpart ($g'=15.5$). However, we did not detect any significant periodic flux modulation in the GROND data, whose cadence did not allow us to homogeneously sample the period space. This is true also for the GROND counterparts to {\it Source \#2} and {\it Source \#6}. As a matter of fact, after inspecting the GROND images we found that the latter is actually a galaxy.

{\it Source \#1} is the brightest of the X-ray sources detected within the 3FGL error circle. Its unabsorbed X-ray flux in the 0.3--10 keV energy range is $F_{\rm X}=7.29^{+0.56}_{-0.54}\times10^{-14}$ erg cm$^{-2}$ s$^{-1}$, assuming, as usual, a PL model with N$_{\rm H}$ fixed to the integrated Galactic value ($N_{\rm H}=4\times10^{20}$ cm$^{-2}$) and photon index $\Gamma_{\rm X}=2.63^{+0.12}_{-0.11}$. We detected neither a significant periodic modulation nor a clear short-term variability in {\it Source \#1}, and this was true for the other \xmm\ sources detected in the 3FGL error ellipse. For the 3FGL\, J1119.9$-$2204 $\gamma$-ray flux $F_{\gamma} = (1.68 \pm 0.10) \times 10^{-11}$ erg cm$^{-2}$ s$^{-1}$ (Acero et al.\ 2015), the X-ray flux of {\it Source \#1} would yield $F_{\gamma}/F_{\rm X}\sim230$. {\it Source \#1} was also the target of the \xmm/pn observations carried out in fast-timing mode. To perform the periodicity analysis on {\it Source \#1}, we used these data, which are characterised by a better timing resolution (0.03 ms) with respect to the MOS detectors (2.6 s), to search for X-ray pulsations and verify whether it was a suitable X-ray counterpart to 3FGL\, J1119.9$-$2204. Besides standard reduction and background screening, we cleaned the pn data by the typical peculiar flaring background component produced by the interactions of charged particles with the CCD (see Burwitz et al.\ 2004) following the same procedure as adopted by De Luca et al.\ (2005). We extracted the source photons using a 8 pixel wide strip (4\farcs1 pixel size) containing 95\% of the source counts and the background photons from two stripes (4 and 4 pixels wide) away from the source region. All photon arrival times were converted to the Solar system Barycentric Dynamical Time (TBD) with the SAS task {\tt barycen} and we used the FTOOL {\tt powspec} to search for periodic signal modulations. However, we did not detect any significant periodic flux modulation down to a period of $\sim1$ ms. This, however, would only rule out the possibility that {\it Source \#1} is an isolated MSP, unless the pulsed fraction is too low for pulsations to be detected in the \xmm\ observation or the X-ray emission is not pulsed because of an unfavourable viewing/beaming geometry. Unfortunately, the available X-ray data do not allow  simultaneous fitting of both a spin and an orbital period. Therefore, we cannot firmly rule out that {\it Source \#1} is, indeed, a MSP in a binary system and, as such, the likely counterpart to 3FGL\, J1119.9$-$2204.

Other X-ray sources detected close to the 3FGL error ellipse are characterised by a $\gamma$--to--X-ray flux ratio $\ga1000$, except for {\it Source \#5} ($F_{\gamma}/F_{\rm X}\sim200$). The 3$\sigma$ detection limit for the \xmm\ observation is $4.1\times10^{-15}$ erg cm$^{-2}$ s$^{-1}$, after combining data from both MOS detectors.

\subsubsection{3FGL\, J1539.2$-$3324}

Only \sw\ observations (84 ks total integration time) of the 3FGL\, J1539.2$-$3324 field have been obtained so far. We detected no candidate X-ray counterpart to the LAT source close to the 3FGL error box (see also Hui et al.\ 2015). The computed $3 \sigma$ sensitivity limit in the 0.3--10 keV energy range is $8.5 \times 10^{-15}$ erg cm$^{-2}$ s$^{-1}$. Its $\gamma$-ray flux $F_{\gamma} = (1.15 \pm 0.10) \times 10^{-11}$ erg cm$^{-2}$ s$^{-1}$ (Acero et al.\ 2015) yields $F_{\gamma}/F_{\rm X} \ga 1350$ for 3FGL\, J1539.2$-$3324. We observed the field of 3FGL\, J1539.2$-$3324 with GROND but no optically variable source has been detected in the 3FGL error ellipse through an automated variability search. 

\subsubsection{3FGL\, J1625.1$-$0021}

The 3FGL\, J1625.1$-$0021 field was observed by both \sw\ (9 ks) and \xmm\ (26 ks). In the \xmm\ observation (OBSID 0672990401) we found three sources close to the 3FGL error ellipse (Fig.~\ref{XrayImages}). We identified the brightest X-ray source ({\it Source \#1}) with source J1625\_X1 of Hui et al.\ (2015). Its unabsorbed X-ray flux in 0.3--10 keV energy range is $F_X=2.22^{+0.40}_{-0.34}\times10^{-14}$ erg cm$^{-2}$ s$^{-1}$ assuming a PL model with hydrogen column density fixed to the integrated Galactic value (N$_{\rm H}=5.8\times10^{20}$ cm$^{-2}$) and photon index $\Gamma_X=3.19^{+0.26}_{-0.25}$. Alternatively, the X-ray spectrum can be fitted by an absorbed BB with fixed N$_H$ and temperature $kT=0.18\pm0.02$ keV. Our spectral analysis is in agreement with that of Hui et al.\ (2015). {\it Source \#2} is the faintest source, characterised by an unabsorbed X-ray flux $F_{\rm X}=6.4^{+0.40}_{-0.39}\times10^{-15}$ erg cm$^{-2}$ s$^{-1}$, obtained by a PL model with fixed column density and photon index ($\Gamma_{\rm X}=2$). {\it Source \#3} is the brightest source detected closest to the edge of the 3FGL error ellipse. Assuming a PL model with fixed column density and $\Gamma_{\rm X}=1.53^{+0.21}_{-0.20}$, the unabsorbed X-ray flux is $F_{\rm X}=2.94^{+0.73}_{-0.53}\times10^{-14}$ erg cm$^{-2}$ s$^{-1}$. The two brightest X-ray sources, {\it Source \#1} and {\it Source \#3}, would have an $F_{\gamma}/F_{\rm X}$ $\sim825$ and $\sim 622$, respectively, assuming the 3FGL\, J1625.1$-$0021 $\gamma$-ray flux $F_{\gamma} = (1.83 \pm 0.12) \times 10^{-11}$ erg cm$^{-2}$ s$^{-1}$ (Acero et al.\ 2015), whereas the $\gamma$--to--X-ray flux ratio for the fainter X-ray source ({\it Source \#2}) would be higher, $\sim2900$. We did not detect either a significant periodic modulation nor a clear short-term variability for the three plausible X-ray counterparts. The 3$\sigma$ point source detection limit for the \xmm\ observation is $6.2\times10^{-15}$ erg cm$^{-2}$ s $^{-1}$.

We detected plausible Catalina counterparts to {\it Source \#2} and {\it Source \#3}, which are CSS\, J162516.0$-$001945/MLS\, J162515.8$-$001944 (V$\sim$20.07) and MLS\, J162509.5$-$002051 (V$\sim22.06$), respectively. However, neither of these two sources shows a significant periodic optical flux modulation. Regardless of an association with an X-ray source, we exploited the fact that the field is covered by the Catalina Surveys Periodic Variable Star Catalogue (Drake et al.\ 2014) to search for variable optical sources. However, we found none. The closest variable Catalina source is CSS\, J162450.6$-$002135 (V=16.48) with a period of 0.628 d, at a distance of 0.09$^{\circ}$ from the centre of the 3FGL error ellipse (r95=0.04$^{\circ}$). We could not find variable optical/infrared sources to the three \xmm\ sources in the GROND data either. 

\subsubsection{3FGL\, J2112.5$-$3044}

The field of 3FGL\, J2112.5$-$3044 has been observed by both \sw\ (3.8 ks) and \xmm\ (33.8 ks). In the \xmm\ observation (OBSID 0672990201), we detected three X-ray sources in the 3FGL error ellipse (see Fig.~\ref{XrayImages}). All spectra were fitted using a PL model with column density fixed to the the integrated Galactic N$_{\rm H}$, that is $\sim6.9\times10^{20}$ cm$^{-2}$. {\it Source \#1} is the brightest X-ray source, with an unabsorbed X-ray flux in the 0.3--10 keV energy range of $F_{\rm X}=1.41^{+0.39}_{-0.22}\times10^{-14}$ erg cm$^{-2}$ s$^{-1}$ and $\Gamma_X=2.59^{+0.29}_{-0.28}$. {\it Source \#2} and {\it Source \#3} are fainter, the PL model fit with fixed N$_{\rm H}$ and $\Gamma_{\rm X}\sim$1.91 and $\sim$2.16, respectively, yields an unabsorbed X-ray flux of $F_{\rm X}=6.6^{+4.9}_{-2.4}\times10^{-15}$ erg cm${-2}$ s$^{-1}$ and $F_{\rm X}=7.9^{+4.5}_{-2.1}\times10^{-15}$ erg cm${-2}$ s$^{-1}$. The $F_{\gamma}/F_{\rm X}$ for three X-ray sources would be between $\sim1300$ and $\sim3000$, for $F_{\gamma} = (1.90 \pm 0.14) \times 10^{-11}$ erg cm$^{-2}$ s$^{-1}$ (Acero et al.\ 2015). We did not detect either a significant periodic modulation nor a clear short-term variability for the three possible X-ray counterparts. None of them is associated with a Catalina counterpart, whereas the field has not been observed by GROND. The computed $3 \sigma$ detection limit of the \xmm\ observation in the 0.3--10 keV energy range is $3.6\times 10^{-15}$ erg cm$^{-2}$ s$^{-1}$. 

\subsubsection{Long-term Variability}

Since for most of our {\em Fermi}-LAT sources we have both \sw\ and \xmm\ observations, we used these data to search for possible long-term variability for the X-ray sources detected within/close to the 3FGL error ellipses, which might spot possible transitional RB candidates. The transition from a rotation-powered to an accretion-powered regime, due to the appearance of an accretion disk, yields a variation of the X-ray flux by a factor of $\sim$10 (e.g., Bogdanov et al.\ 2015).

For each \sw\ observation we extracted the source events from a circular region with radius of 25\arcsec centred on the position detected by the more sensitive instruments on board of \xmm, while to evaluate the background we extracted events from a source-free region of 130\arcsec-radius. We generated the ancillary response files with the {\it xrtmkarf} task accounting for different extraction regions, vignetting and point-spread function corrections and we used the latest available spectral redistribution matrix (v014). We estimated the source flux by fitting in the 0.3--10 keV band to each spectrum a PL model, with hydrogen column density and X-ray photon index fixed to the best-fit values obtained by \xmm, see Table \ref{tab2}. In those cases in which the count rate in an observation was compatible with zero, we set an upper limit at the 3$\sigma$ level.

\sw\ observed three times 3FGL\, J0802.3$-$5610 in 2012, 27 times 3FGL\, 1119.9$-$2204 from 2010 to 2013, twice 3FGL\, J1625.1$-$0021 from 2010 to 2011, and 6 times 3FGL\, J2112.5$-$3044 from 2013 to 2014. The flux values measured for each X-ray source with \sw\ and \xmm\ do not display any significant long-term variability. Though almost all \sw\ observations provided 3$\sigma$ upper limits on the flux because they are often very short (of the order of few ks) and the X-ray sources detected are very faint, these values are in agreement with those obtained by more sensitive instruments.

\subsection{Candidates with a recent pulsar identification}\label{sec:3.3}

\subsubsection{3FGL\, J1035.7$-$6720}

The field has been observed in X-rays both by \sw\ and \xmm\ (OBSID 0692830206) for a total integration time of 43 ks and 25 ks, respectively. A candidate X-ray counterpart to the LAT source has been found in the \xmm\ data by Saz-Parkinson et al.\ (2016). The source X-ray spectrum is fit by a PL with photon index $\Gamma_{\rm X}=2.91^{+0.46}_{-0.40}$, for a fixed $N_{\rm H} = 2 \times 10^{21}$ cm$^{-2}$, with an unabsorbed 0.3--10 keV flux $F_{\rm X} = 3.06^{+0.97}_{-0.50} \times 10^{-14}$ erg cm$^{-2}$ s$^{-1}$. For the 3FGL\, J1035.7$-$6720 $\gamma$-ray flux $F_{\gamma} = (2.59 \pm 0.14) \times 10^{-11}$ erg cm$^{-2}$ s$^{-1}$ (Acero et al.\ 2015) the source X-ray flux corresponds to $F_{\gamma}/F_{\rm X} \sim 850$. The field of 3FGL\, J1035.7$-$6720 is not covered by the Catalina Sky Survey but has been observed with GROND. No optical/infrared counterpart, however, has been detected at the X-ray source position and no variable optical/infrared object has been found in the rest of the 3FGL error ellipse through an automated variability search.

While this paper was close to being finalised, 3FGL\, J1035.7$-$6720 was detected as a $\gamma$-ray pulsar by the {\it Einstein@Home} survey (Clark et al.\ 2017), although it is not neither known whether it is isolated or binary nor whether it is an MSP or a young pulsar. This source is under investigation by these authors because of its peculiar timing properties and a more detailed account will be given in a forthcoming publication (Clark et al., in preparation).

\subsubsection{3FGL\, J1630.2+3733}

The {\em Fermi}-LAT source 3FGL\, J1630.2+3733 has been recently identified by the {\em Fermi} Pulsar Search Consortium to be a binary MSP (PSR\, J1630+3734), with a spin period of P$_{\rm s}$ = 3.32 ms and an orbital period P$_{\rm orb}$ = 12.5 d (Ray et al.\ 2012; Sanpa-Arsa et al., in prep.). The pulsar coordinates are not published with a degree of accuracy good enough to identify its counterpart at other wavelengths based upon positional coincidence. Nonetheless, we searched for a possible X-ray counterpart to PSR\, J1630+3734 in the \sw\ data, which consists of only a snapshot 4.8 ks exposure. We did not find X-ray sources close to the 95\% 3FGL error box. The computed $3 \sigma$ sensitivity limit in the 0.3--10 keV energy range is $6.1 \times 10^{-14}$ erg cm$^{-2}$ s$^{-1}$. This represents the first constraint on the pulsar X-ray flux obtained so far. No observations with other X-ray satellites have been obtained for this source. Therefore, we can set a limit of $F_{\gamma}/F_{\rm X} \ga110$ for $F_{\gamma} = (6.86 \pm 0.10) \times 10^{-12}$ erg cm$^{-2}$ s$^{-1}$ (Acero et al.\ 2015).

Since the \sw\ observation is relatively shallow, the X-ray counterpart to the MSP might be below the detection limit. Therefore, like we did for the 3FGL\, J1625.1$-$0021 field, we searched for variable sources in the Catalina Surveys Periodic Variable Star Catalogue (Drake et al.\ 2014) which might not be associated with a bright X-ray source. However, also in this case we found none. The field has been observed also by the \sw/UVOT but only 6 U-band exposures (4.8 ks total) are available, not enough to pinpoint a candidate companion star to the MSP based upon optical variability. Therefore, the companion star to PSR\, J1630+3734 remains unidentified. 

\subsubsection{3FGL\, J1744.1$-$7619}

The field has been observed both by \sw\ and \xmm\ (OBSID 0692830101) for a total integration time of 6.9 ks and 25.9 ks, respectively (see also Hui et al.\ 2015). A candidate X-ray counterpart to the LAT source has been found in the \xmm\ data by Saz-Parkinson et al.\ (2016), identified as source J1744\_X1 of Hui et al.\ (2015). The source X-ray spectrum is fit by a PL with photon index $\Gamma_{\rm X}=2.71^{+0.40}_{-0.39}$, for a fixed $N_{\rm H} = 8 \times 10^{20}$ cm$^{-2}$, with an unabsorbed 0.3--10 keV flux $F_{\rm X} = 1.92^{+0.59}_{-0.39} \times 10^{-14}$ erg cm$^{-2}$ s$^{-1}$ (Saz Parkinson et al.\ 2016). For the 3FGL\, J1744.1$-$7619 $\gamma$-ray flux $F_{\gamma} = (2.25 \pm 0.13) \times 10^{-11}$ erg cm$^{-2}$ s$^{-1}$ (Acero et al.\ 2015) this corresponds to $F_{\gamma}/F_{\rm X} \sim 1170$. The field of 3FGL\, J1744.1$-$7619 is only covered by the Catalina Sky Survey but no optical counterpart is found at the X-ray source position. Like in the case of 3FGL\, J1035.7$-$6720, 3FGL\, J1744.1$-$7619 has been recently detected as a $\gamma$-ray pulsar by the {\it Einstein@Home} survey (Clark et al.\ 2017) and it is also under investigation.

\section{Summary and Discussion}

As part of our survey of candidate MSPs we identified a new candidate RB system (3FGL\, J2039.6$-$5618; Salvetti et al.\ 2015) through correlated optical/X-ray flux modulations, we independently confirmed  3FGL\, J0523.3$-$2528 (Strader et al.\ 2014) and 3FGL\, J1653.6$-$0158 (Romani et al.\ 2014) as binary MSP candidates, we found a new likely binary MSP candidate through the detection of a periodic optical modulation (3FGL\, J0744.1$-$2523). For 3FGL\, J0802.3$-$5610 we found only a marginal evidence of optical periodicity that may be affected by the presence of a strong daily periodicity associated with the cadence of the Catalina observations. For this reason, follow-up optical observations are required to confirm such periodicity and identify 3FGL\, J0802.3$-$5610 as a new binary MSP. For both 3FGL\, J1539.2$-$3324 and the recently identified binary MSP 3FGL\, J1630.2+3733$\equiv$PSR\, J1630+3734 (Sanpa-Arsa et al.\, in prep.), we could find neither a candidate X-ray counterpart nor a variable optical counterpart candidate within the 3FGL error ellipse. For the other sources (3FGL\, J1035.7$-$6720, J1119.9$-$2204, J1625.1$-$0021, J1744.1$-$7619, J2112.5$-$3044) we detected up to several candidate X-ray counterparts but we could not find an association with a periodically-modulated optical counterpart or with any optical counterpart at all. Therefore, like for 3FGL\, J1539.2$-$3324, their proposed identification as binary MSPs cannot be confirmed yet, although both 3FGL\, J1035.7$-$6720 and 3FGL\, J1744.1$-$7619 have now been reported to be pulsars of still unspecified type (Clark et al.\ 2017). For the others, deeper X-ray and optical observations are needed to ascertain their nature.

For 3FGL\, J2039.6$-$5618, we exploited archival {\em WISE} data (Mignani et al.\ 2016) to obtain a better characterisation of its multi-band spectrum with respect to Salvetti et al.\ (2015). Interestingly, its multi-band UV--to--mid-IR spectrum shows a bump in the near/mid-infrared emission, that can be explained by the presence of two BB components (Sectn.~\ref{sect_J2039}). Since modulations are seen in the GROND optical bands (Salvetti et al.\ 2015) the hotter BB component is likely associated with the emission from the tidally-distorted companion star. Since the same modulations are also seen in the JHK GROND light curves, the emission from the companion star must contribute in the near-IR as well. The colder BB, however, cannot be entirely associated with emission from the star, and its origin is unclear. As we suggested in Mignani et al.\ (2016), it might be associated with emission from cold intra/extra-binary material, perhaps associated with the ablated gas from the companion or a residual of an accretion disk, or both. Interestingly, the size of the emission region is a few times that of the star Roche Lobe, which would corroborate the latter hypothesis. Setting tighter constraints on the absence of periodicity in the multi-epoch {\em WISE} data (Sectn.~\ref{sect_J2039}) of the source would be important to confirm the hypothesis that the mid-IR emission from the system indeed comes from diffuse gas and not from the surface of the MSP companion.

For 3FGL\, J0744.1$-$2523, the detection of a 0.115 d periodicity in its candidate optical counterpart would exclude the association with a binary system consisting of two classical stars (W UMa or $\beta$ Lyr binaries) because their orbital period is greater than 0.22 days (Geske et al.\ 2006). The association with a cataclysmic variable (CV) would also be very unlikely because the period is within the 2--3 hours period gap of the CV period distribution (Southworth et al.\ 2015). We note that, in principle, a period at 0.230 d, i.e. two times the value that we found, cannot be excluded a priori. By folding the GROND data at the 0.230 d period we obtain a light curve with two peaks, with the first one only hinted at owing to the lack of data in the 0--0.25 phase range. Therefore, we cannot determine whether the two peaks are asymmetric, which would speak in favour of a genuine periodicity, or not. We can only say that, within the uncertainties, the shape of the first peak seems compatible with the second one. A 0.230 d period would make alternative scenarios for 3FGL\, J0744.1$-$2523 more difficult to rule out. Only the detection of the source in the X-rays, or even very constraining limits on its X-ray emission, together with an optical spectroscopy analysis aimed at measuring the radial velocity curve of the star will allow us to determine which period is the real one and which is an alias and, thus, to firmly discriminate between different scenarios.

We note that it is possible that some of the $\gamma$-ray sources that have a candidate X-ray identification, but are not associated with an optical counterpart (i.e. 3FGL\, J1035.7$-$6720, J1744.1$-$7619, J2112.5$-$3044), are isolated rather than binary MSPs. However, this hypothesis cannot be verified with the current X-ray/optical data. Indeed, the low X-ray flux of the candidate X-ray counterpart to 3FGL\, J1035.7$-$6720 makes the constraint on the X-ray--to--optical flux ratio $F_{\rm X}/F_{\rm opt}$ obtained from the GROND observations not compelling enough to identify this $\gamma$-ray source as an isolated pulsar/MSP. For instance, for an $F_{\rm X}/F_{\rm opt}\ga 1000$, typical of isolated pulsars (e.g. Mignani 2011), the flux of the candidate X-ray counterpart to 3FGL\, J1035.7$-$6720 ($F_{\rm X} = 3.06^{+0.97}_{-0.50} \times 10^{-14}$ erg cm$^{-2}$ s$^{-1}$) would require an upper limit to its optical flux corresponding to a magnitude $g'\sim28.5$, clearly unattainable with GROND. This is even more so for the candidate X-ray counterparts to 3FGL\, J1744.1$-$7619 and 3FGL\, J2112.5$-$3044, for which the optical flux upper limits that can be obtained from the Catalina data are at least 2--3 magnitudes shallower than the GROND ones. Deep observations with 10-m class telescopes are, thus, required to obtain constraining $F_{\rm X}/F_{\rm opt}$ values for all these sources.

\begin{figure*}
{\includegraphics[width=8cm,angle=0,clip=true]{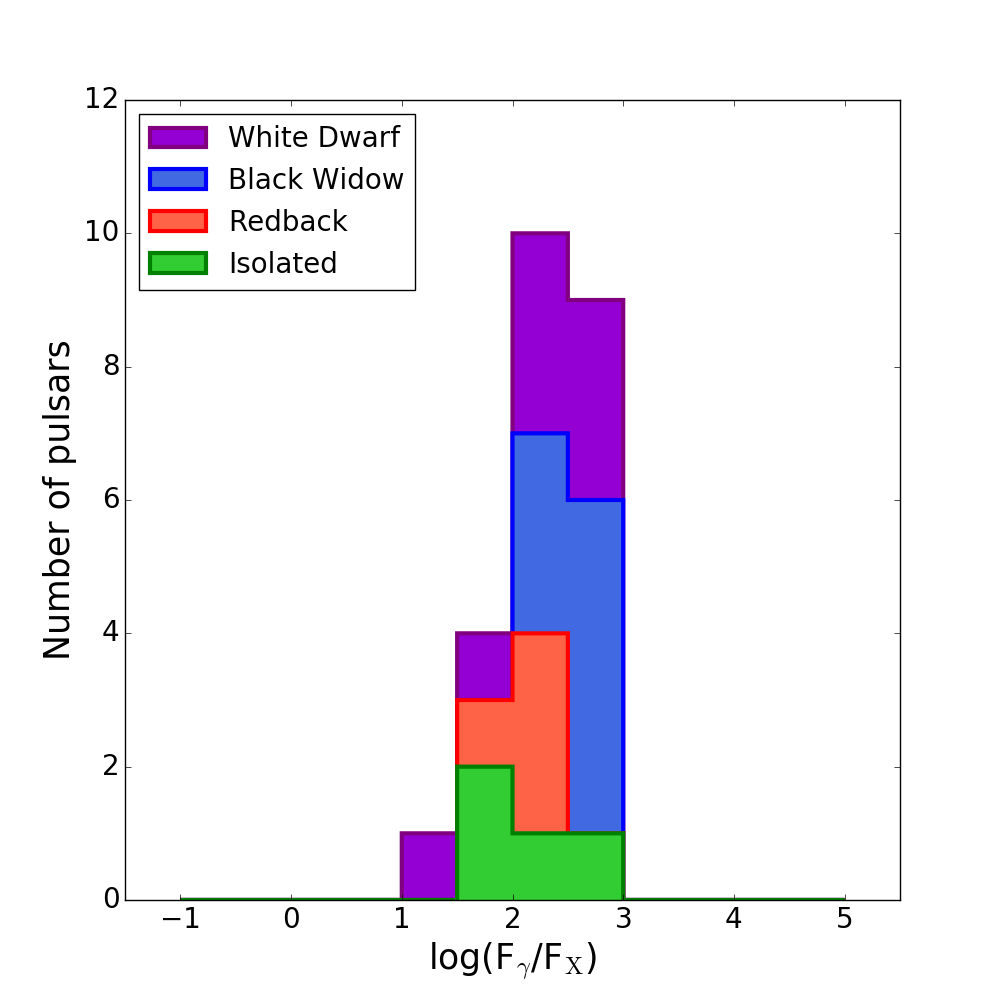}}
{\includegraphics[width=8cm,angle=0,clip=true]{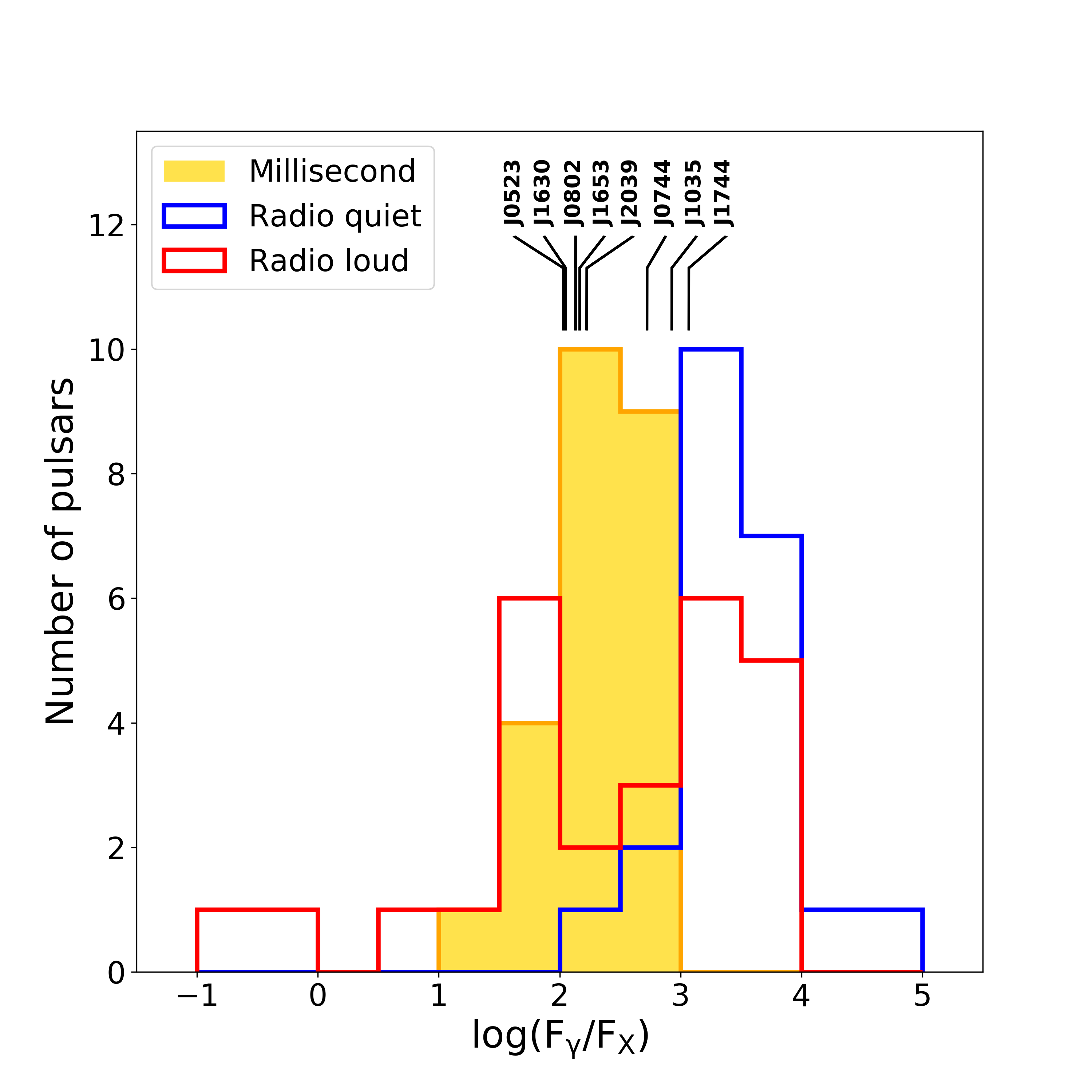}}
\caption{Histograms of the $F_{\gamma}/F_{\rm X}$ ratio for different classes of MSPs detected in $\gamma$-rays by the {\em Fermi}/LAT (left) and for all $\gamma$-ray pulsars (right). In the right panel, the vertical ticks, labelled with the 3FGL source names, show the $F_{\gamma}/F_{\rm X}$ values for the high-confidence candidate RBs in Sectn.~\ref{sec:3.1} and for the recently identified pulsars (Sectn.~\ref{sec:3.3}), together with the values for the candidate binary MSP 3FGL\, J0744.1$-$2523  and 3FGL\, J0802.3$-$5610
(Sectn.~\ref{sec:3.2}).
For both 3FGL\, J0744.1$-$2523 and 3FGL\, J1630.2+3733 the plotted values correspond to lower limits.}
\label{histo}
\end{figure*}

We exploited the value of the $F_{\gamma}/F_{\rm X}$ ratio as a diagnostic to determine whether some of the X-ray sources detected in the 3FGL error ellipses could be plausible MSP candidates.
The distribution of this ratio for different classes of MSPs (isolated, RBs, BWs, binary with WD companions) detected as $\gamma$-ray pulsars is shown in Figure \ref{histo} (left) compared
to that of the young RL and RQ $\gamma$-ray pulsars (Figure \ref{histo}, right), updated from Marelli et al.\ (2015). We computed the histograms for the MSPs using the same criteria as in Marelli et al.\ (2015) for young RL and RQ pulsars.
We selected a starting sample of $\gamma$-ray MSPs from the Public List of LAT-Detected Gamma-Ray Pulsars. Then, from the literature we selected those MSPs with a clear, non-thermal X-ray spectrum, either associated with emission from the pulsar magnetosphere and/or an intra-binary shock in the case of BW and RB systems (Roberts 2013), and we assumed the non-thermal X-ray flux as a reference. We warn here that for BWs and RBs the emission from the intra-binary shock cannot be easily disentangled from that of the MSP magnetosphere, unless the detection of X-ray pulsations makes it possible to carry out phase-resolved spectroscopy to separate the pulsed component, so that some uncertainty on the latter component is expected. For MSPs with a composite spectrum, i.e. featuring also thermal X-ray emission from hot polar caps, we considered only the non-thermal component. Finally, we filtered out transitional MSPs, whose X-ray emission can, at times, result from matter accretion from the companion star. We used the classifications reported in the on-line Millisecond Pulsar Catalogue compiled by A. Patruno\footnote{https://apatruno.wordpress.com/about/millisecond-pulsar-catalogue/} as a reference to select the different classes of MSPs.

As we see, there is no substantial difference in the values of the $F_{\gamma}/F_{\rm X}$ for different classes of MSPs, although possible differences might be hidden by the still relatively
small samples and be uncovered through future detections. The peak of the $F_{\gamma}/F_{\rm X}$ distribution for MSPs is somewhere in between the two peaks of the corresponding distribution for RL pulsars. 
Differences in distribution of $F_{\gamma}/F_{\rm X}$ values for RQ and RL pulsars are usually interpreted in differences in geometries and/or emission zones of X- and $\gamma$-rays in the magnetosphere.
MSPs have a more compact magnetosphere than young pulsars, so that we can argue that the emission zones for different wavelengths should be nearer to each other in MSPs and, therefore, less dependent on the pulsar geometry.
In a way, such a sharp distribution in the $F_{\gamma}/F_{\rm X}$ for MSPs, with a mean value consistent with that of the RL pulsars, is expected.
With such a rapidly increasing sample of new MSPs, new observations and theoretical models for MSPs magnetosphere are fundamental to study the mechanisms involved in pulsars' emission.  

Interestingly, the $F_{\gamma}/F_{\rm X}$ ratios of the candidate RBs 3FGL\, J0523.3$-$2528, 3FGL\, J1653.6$-$0158, and 3FGL\, J2039.6$-$5618, and that of the new binary MSP 3FGL\, J1630.2+3733$\equiv$PSR\, J1630+3734 are all in the range $\sim 100$--170, close to that of the known MSPs identified in $\gamma$-rays. The $F_{\gamma}/F_{\rm X}$ ratios for 3FGL\, J1035.7$-$6720 and 3FGL\, J1744.1$-$7619, which have been recently announced to be $\gamma$-ray pulsars (Clark et al.\ 2017), are $\sim 850$ and $\sim 1170$, respectively. These values would still be consistent with an identification as MSPs or as isolated RL/RQ $\gamma$-ray pulsars.

For the remaining six $\gamma$-ray sources (Table~\ref{tab2}), the $F_{\gamma}/F_{\rm X}$ ratio computed for their possible X-ray counterparts covers a wide range of values, from $\sim$300 up to $\sim$5000, which overlaps the distribution for all $\gamma$-ray pulsars (Figure \ref{histo}, right). However, we note that in the case of 3FGL\, J0802.3$-$5610 the $F_{\gamma}/F_{\rm X}$ ratio for {\it Source \# 5}, which is the most likely X-ray counterpart to the $\gamma$-ray source (Sectn.~\ref{sect_J0802}), is $135^{+81}_{-52}$. This value is consistent with those of the candidate RBs 3FGL\, J0523.3$-$2528, 3FGL\, J1653.6$-$0158, and 3FGL\, J2039.6$-$5618 and the binary MSP 3FGL\, J1630.2+3733. This is also true for {\it Source \#1} ($F_{\gamma}/F_{\rm X}=230^{+22}_{-22}$), which is the most likely X-ray counterpart to 3FGL\, J1119.9$-$2204 (Sectn.~\ref{sect_J1119}). For 3FGL\, J0744.1$-$2523, the available X-ray detection limit points at a higher $F_{\gamma}/F_{\rm X}$ ($\ga 525$) but still compatible with the distribution for MSP. Therefore, the MSP identification for these three sources appears, at least, plausible.

\section{Conclusions}

We used observations in the X-rays (\xmm, \chan, \sw) and in the optical (Catalina, GROND) of twelve 3FGL $\gamma$-ray sources that were originally {\em unassociated} when this project started and classified as likely MSP candidates based upon statistical classification techniques (Salvetti 2016; Saz-Parkinson et al.\ 2016). In the course of this project, four of these sources were identified as either binary MSPs from the detection of radio/$\gamma$-ray pulsations (Sanpa-Arsa et al., in prep) or candidate RBs from the detection of a periodic optical modulation (Strader et al.\ 2014; Romani et al.\ 2014), including 3FGL\, J2039.6$-$5618 (Salvetti et al.\ 2015; Romani 2015), whereas another two were announced to be pulsar of yet unknown class (Clark et al.\ 2017). This speaks very much in support of our classification technique as well as of the multi-wavelength approach employed to investigate binary MSP candidates, which has encouraged follow-up observations of other well-ranked candidates with optical/X-ray facilities, now in progress.

 For the known RB candidate 3FGL\, J2039.6$-$5618 we might have found evidence of the presence of intra-binary gas from its peculiar double-BB optical--to--mid-IR spectrum, a feature now being searched for in other objects of this sort, which might track the ablation process of the companion star or the past presence of an accretion disk, possibly providing a new diagnostic to pinpoint transitional MSP candidates. We also found a new binary MSP candidate in 3FGL\, J0744.1$-$2523, owing to the detection of a clear $\sim 0.115$ day periodicity in its putative optical counterpart. Like for 3FGL\, J2039.6$-$5618, the double-BB spectrum hints at the presence of intra-binary gas. Radial velocity measurements for the optical counterparts to both sources, now in progress, will provide the orbital parameters of the binary systems and secure their identification as binary MSPs. For a third $\gamma$-ray source, 3FGL\, J0802.3$-$5610, we have found only a marginal evidence of periodic optical modulations at a period expected for a compact binary system, which singles out this source for deeper investigations both in the optical and in the X-rays. For the remaining sources, we cannot confidently rule out their identification as binary MSPs without more compelling observational support which can only come from dedicated observations.

\section*{Acknowledgments}

We thank the anonymous referee for his/her very helpful comments to our manuscript. DS, AB and MM acknowledge support through EXTraS, funded from the European Union's Seventh Framework Programme for research, technological development and demonstration under grant agreement no 607452. RPM acknowledges financial support from the project TECHE.it. CRA 1.05.06.04.01 cap 1.05.08 for the project ``Studio multi-lunghezze d'onda da stelle di neutroni con particolare riguardo alla emissione di altissima energia''. The CSS survey is funded by the National Aeronautics and Space Administration under Grant No. NNG05GF22G issued through the Science Mission Directorate Near-Earth Objects Observations Program. The CRTS survey is supported by the U.S. National Science Foundation under grants AST-0909182. Part of the funding for GROND (both hardware as well as personnel) was generously granted from the Leibniz-Prize to Prof. G. Hasinger (DFG grant HA 1850/28-1).

\label{lastpage}

\end{document}